\documentclass[journal,comsoc]{IEEEtran}

\ifCLASSINFOpdf
  \usepackage[pdftex]{graphicx}
  \DeclareGraphicsExtensions{.pdf,.jpeg,.png}
\else
  \usepackage[dvips]{graphicx}
\fi

\usepackage[algoruled,linesnumbered]{algorithm2e}
\usepackage{caption}
\usepackage{subcaption}
\usepackage{multirow}
\usepackage{tabularx}
\usepackage{hhline}
\usepackage{hyperref}
\usepackage[cmex10]{amsmath}
\usepackage{amsthm,amssymb}
\usepackage[cmintegrals]{newtxmath}
\usepackage{fancyhdr}
\usepackage{epstopdf}

\hypersetup{pdfauthor={Author's name},%
pdftitle={Document Title},%
pagebackref=true,%
colorlinks=true,%
pdftex
}

\usepackage{xr}
{\endproof}

\begin{document}

\title{Cooperative Data Offload in Opportunistic Networks: From Mobile Devices to Infrastructure}

        
\author{Zongqing~Lu,~Xiao~Sun,~and~Thomas~La~Porta


\thanks{Z. Lu, X. Sun and T. La Porta are with the Department of Computer Science and Engineering, The Pennsylvania State University, University Park, PA 16802, US. E-mail: \{zongqing, xxs118, tlp\}@cse.psu.edu.}
}

\maketitle

\begin{abstract}
Opportunistic mobile networks consisting of intermittently connected mobile devices have been exploited for various applications, such as computational offloading and mitigating cellular traffic load. In contrast to existing work, in this paper, we focus on cooperatively offloading data among mobile devices to maximally improve the probability of data delivery from a mobile device to intermittently connected infrastructure within a given time constraint, which is referred to as the \textit{cooperative offloading} problem. Unfortunately, the estimation of data delivery probability over an opportunistic path is difficult and cooperative offloading is NP-hard. To this end, we first propose a probabilistic framework that provides the estimation of such probability. Based on the proposed probabilistic framework, we design a heuristic algorithm to solve cooperative offloading at a low computation cost. Due to the lack of global information, a distributed algorithm is further proposed. The performance of the proposed approaches is evaluated based on both synthetic networks and real traces. Experimental results show that the probabilistic framework can accurately estimate the data delivery probability, cooperative offloading greatly improves the delivery probability, the heuristic algorithm approximates the optimum, and the performance of both the heuristic algorithm and distributed algorithm outperforms other approaches.
\end{abstract}


\IEEEpeerreviewmaketitle

\section{Introduction}
\IEEEPARstart{O}{pportunistic} mobile networks consist of mobile devices which are equipped with short-range radios (e.g. Bluetooth, WiFi). Mobile devices intermittently contact each other without the support of infrastructure when they are within range of each other. Most research work on such networks focuses on data forwarding \cite{spyropoulos2005spray}\cite{erramilli2008delegation} and data caching \cite{gilbert2011peer}\cite{zhuo2011contact}. Other research work focuses on exploring opportunistic mobile networks for various applications. Shi \emph{et al.} \cite{shi2012serendipity} proposed to enable remote computing among mobile devices so as to speedup computing and conserve energy. Liu \emph{et al.} \cite{liu2013exploring} explored the practical potential of opportunistic networks among smartphones. Lu \emph{et al.} \cite{lu2016networking} built a system to network smartphones opportunistically using WiFi for providing communications in disaster recovery. Han \emph{et al.} \cite{han2012mobile} proposed to migrate the traffic from cellular networks to opportunistic communications between smartphones, coping with the explosive traffic demands and limited capacity provided by cellular networks.

In this paper, we focus on exploiting opportunistic communications for offloading data which is to be transmitted from a mobile device to intermittently connected infrastructure (e.g., a remote server). Specifically, besides directly transmitting data to infrastructure, the mobile device can also transfer data to other devices and then let these devices transmit the data to infrastructure so as to improve the probability of successful data delivery. Unlike mobile cloud computing, which assumes mobile devices are always connected to the cloud via cellular networks and the decision of offloading is mainly based on energy consumption or application execution time \cite{cuervo2010maui}\cite{chun2011clonecloud}\cite{xiang2014ready}\cite{zhang2015collaborative}, we consider the data offloading scenario where mobile devices only have intermittent connections with infrastructure, and mobile devices collaborate to transmit data to infrastructure. 

There are several application scenarios. For example, in vehicular ad hoc networks, instead of deploying more roadside units to increase the coverage, a vehicle can seek help from other vehicles for transmitting data to roadside units when it will not meet roadside units on its current route or it cannot completely transfer the data to roadside units due to their transient contact. Therefore, such a cooperative data offloading scheme that takes advantage of the communication capability of other vehicles can potentially reduce the cost of building more roadside units, while maintaining the desired coverage.
In disaster recovery, due to the very limited coverage and bandwidth of deployed mobile cellular towers, data can be fragmented and sent to multiple users for a better reachability to the mobile cellular towers.  

In opportunistic mobile networks, data replication \cite{spyropoulos2005spray} and data redundancy \cite{jain2005using} are normally employed to improve the probability of successful data delivery by increasing the amount of transmitted data. However, both techniques lead to high bandwidth overhead, which can be severe for large data transfer and costly for cellular networks. Unlike existing work, we focus on offloading data segments to other mobile devices so as to improve the probability of successful data delivery. We rely on an important observation that the probability of successful data delivery over the opportunistic path drops dramatically when the size of transmitted data is large. In such scenarios, offloading data segments can yield a better delivery probability than data replication or data redundancy, in addition to not incurring additional bandwidth overhead. However, maximally improving the delivery probability by data offloading, which is referred to as the \textit{cooperative offloading} problem, turns out to be non-trivial and NP-hard. Although cooperative offloading admits PTAS (Polynomial Time Approximation Scheme), PTAS is still not affordable for mobile devices due to its incurred computation overhead of data delivery probability. 

To deal with this, we design a centralized heuristic algorithm based on the proposed probabilistic framework to solve cooperative offloading. To cope with the lack of global information, we further propose a distributed algorithm. Through extensive simulations on synthetic networks and experiments on real traces, we demonstrate that cooperative offloading can greatly improve the data delivery probability, the heuristic algorithm approximates the optimum, and the performance of both the centralized heuristic algorithm and distributed algorithm outperforms other approaches. The main contributions of this paper can be summarized as follows.

\begin{itemize}
\item We propose a probabilistic framework that provides the estimation of data delivery probability over the opportunistic path, considering both data size and contact duration. To the best of our knowledge, this is the first work that gives such estimation without any restrictions. 
\item We design a heuristic algorithm that performs path allocation and data assignment by carefully considering opportunistic contact probability and path capacity. The algorithm approximates the optimum at a much lower computational cost than PTAS.
\item We design a distributed algorithm which employs criterion assignment, real-time adjustment and assignment update to make data offloading decisions at runtime. Although it only exploits paths no more than two hops for data forwarding, its performance is comparable to that of the heuristic algorithm. 
\end{itemize}

The rest of this paper is organized as follows. Section \ref{sec:RelatedWork} reviews related work and Section \ref{sec:Statement} gives the overview. The probabilistic framework is presented in Section \ref{sec:framework}, followed by the heuristic algorithm and the distributed algorithm in Section \ref{sec:Heuristic} and Section \ref{sec:Distributed}, respectively. Section \ref{sec:Evaluation} evaluates the performance of the proposed approaches and Section \ref{sec:Conclusion} concludes the paper.

\section{Related Work}
\label{sec:RelatedWork}
Mobile opportunistic networks have been studied mainly for data forwarding, where mobile nodes carry and forward messages upon intermittent node contacts. The key problem is how to select appropriate relays such that messages can be forwarded to destinations quickly considering forwarding costs. Many forwarding schemes have been proposed from different perspectives, such as \textit{representative} strategies including \cite{erramilli2007diversity}\cite{erramilli2008delegation}\cite{spyropoulos2008efficient}, \textit{contact capability} based schemes including \cite{yuan2009predict}, and \textit{social concepts} based schemes including\cite{daly2007social}\cite{hossmann2010know}\cite{zhang2013transient}\cite{pietilanen2012dissemination}\cite{lu2014skeleton}\cite{lu2015algorithms}. In addition, forwarding is addressed as a resource allocation problem in \cite{balasubramanian2007dtn} and forwarding capability of nodes under energy constraints is investigated in \cite{banerjee2007energy}. However, most research work on forwarding in mobile opportunistic networks is based on an assumption that nodes can completely transfer a data item during a node contact. Nevertheless, this assumption is not valid in some cases, for example, when node contacts are transient (e.g., in vehicular ad hoc networks) or when data items are large. 

Unlike the existing work, in this paper, we relax this assumption and take contact duration into consideration. The selection of forwarding path(s) for a data item is determined based on the size of the data item; i.e., we choose the path(s) with the maximum probability of successful delivery of the data item by incorporating data fragmentation (i.e., the data item can be fragmented and the segments can be forwarded over different paths). Note that social relations (e.g., community) that are exploited to characterize node contacts cannot provide the estimation of data delivery probability.

The explosive growth of mobile devices (such as smartphones and tablets) has stimulated research on exploring mobile opportunistic networks for various applications. These applications can be generally classified into three categories: mitigating cellular traffic load, offloading computational tasks and information dissemination. The potential of mobile opportunistic networks in terms of reducing cellular traffic load was investigated in \cite{liu2013exploring}. The incentive mechanism was studied in \cite{zhuo2014incentive} and some solutions were proposed in \cite{han2012mobile}\cite{wang2014toss}\cite{sermpezis2014not}\cite{sciancalepore2016offloading}. Computational offloading among intermittently connected mobile devices was investigated in \cite{shi2012serendipity}\cite{li2014can}. Information dissemination in mobile opportunistic networks was considered based on mobility and community in \cite{wang2014tempo} and \cite{lu2014information}, respectively. Unlike the existing work, in this paper, we investigate application scenarios in which mobile devices cooperatively transmit data to intermittently connected infrastructure so as to improve the probability of data delivery.

\begin{figure*}[!t]
\setlength{\abovecaptionskip}{7pt}
\setlength{\belowcaptionskip}{-10pt}
\centering
\begin{minipage}{.31\textwidth}
  \centering
  \includegraphics[width=.55\textwidth]{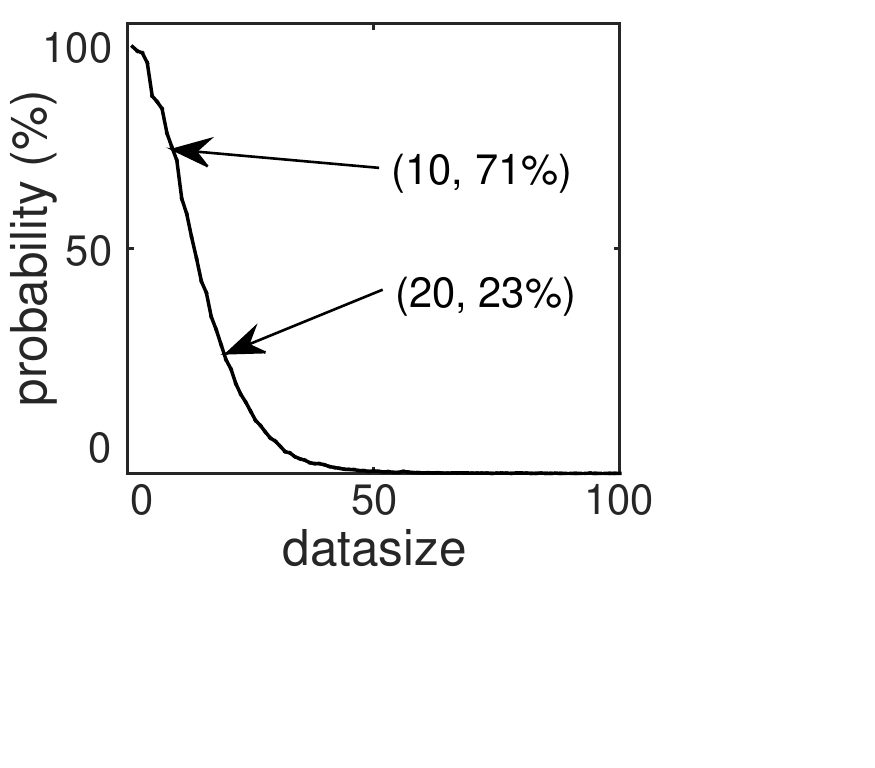}
  \caption{Relation between data size and successful delivery probability for a given pair of path and time constraint.}
  \label{fig:relation}
\end{minipage}
\hspace{0.2cm}
\begin{minipage}{.6\textwidth}
\begin{minipage}{.5\textwidth}
  \centering
  \includegraphics[width=.6\textwidth]{MobileOffloading.15} 
\end{minipage}%
\begin{minipage}{.5\textwidth}
\renewcommand{\arraystretch}{1.2}
\begin{small}
\begin{tabular}{c|cc}
\multicolumn{3}{c}{Successful delivery probability} \\
\hline
\multirow{2}{*}{data size}          & \multicolumn{2}{c}{path}    \\  \cline{2-3}
		  &  $i$   &  $j$    \\  \hline \hline
$S=20$       &  0.23  &  0.23   \\
$S/2$     &  0.71  &  0.71   \\  \hline
\end{tabular}
\end{small}
\end{minipage}
\vspace{0.2cm}
\caption{A simple scenario where node $u$ and $v$ are connected by two paths. Table shows the delivery probabilities of different data sizes over the paths, taken from Fig. \ref{fig:relation}.}
 \label{fig:example}
\end{minipage}
\end{figure*}

\section{Overview}
\label{sec:Statement}

\subsection{Network Model}
Consider a network, where mobile nodes opportunistically contact each other and each node may intermittently connect with infrastructure. Then, mobile nodes and infrastructure together form an opportunistic mobile network. To simplify the notation and the presentation of the paper, infrastructure is seen as a special node, and intermittent connections between nodes and infrastructure are seen as node contacts. In the network, for two nodes, an edge exists between them only if they opportunistically contact each other. With regard to data offloading, each node can exploit other nodes for data transmission. Specifically, in addition to transmitting a data item to the infrastructure along one particular path, the node can also select multiple paths to the infrastructure, and each path can be used to carry part of the data item.


\subsection{Basic Idea}
In an opportunistic mobile network, given the pattern of contact frequency and contact duration between nodes, we first investigate the relation between data size and successful delivery probability for given path and time constraint. Intuitively, since the size of data that can be transmitted during a node contact is restricted by the contact duration, data with small size can be easily transmitted over the opportunistic path; data with large size may require multiple node contacts and thus not be completely transmitted. As shown in Figure \ref{fig:relation}, which is obtained for a given pair of path and time constraint (taken from the experiments in Section \ref{sec:Evaluation}), the probability of successful delivery exponentially decreases when the data size increases. In other words, the delivery probability can exponentially increase if less data is transmitted over the opportunistic path. Based on this important observation, it is concluded that a node may obtain a better delivery probability of a data item if it selects multiple paths to the destination and each path carries a part of the data item. 

To better elaborate on this motivation, let us consider a simple example as shown in Figure \ref{fig:example}, where node $u$ needs to send a data item $S$ (we abuse the notation a little and let $S$ also denote the size of the data item) to $v$. Let us use the real delivery probabilities over an opportunistic path as labeled in Figure \ref{fig:relation}, and $S=20$. Assume paths $i$ and $j$ have the same delivery probability for $S$ and $S/2$. If $u$ transmits $S$ to $v$ over path $i$ or $j$, the delivery probability is 0.23. If each path carries a replication of $S$, the probability is $\sum_{n=1}^2 \binom{2}{n} 0.23^n0.77^{2-n}=0.407$, considering all failures are Bernoulli. Instead, if $u$ sends $S/2$ along each path, the delivery probability is $0.71 \times 0.71 = 0.504$, which is more than two times of the probability when $S$ is sent over individual path. 

It may be surprising that sending data segments along multiple paths can achieve a better delivery probability, even better than sending replications along multiple paths. Although increasingly sending replications over multiple paths may eventually yield better performance, this approach is restricted by available paths and incurs a cost of multiplied network resources. In this example, sending $S$ over four paths between $u$ and $v$ will result in the delivery probability of $\sum_{n=1}^4 \binom{4}{n} 0.23^n0.77^{4-n} > 0.504$, but at the cost of four times of network traffic and thus bandwidth overhead, not to mention requiring four paths between the source and destination. Moreover, applying data redundancy (e.g., erasure coding) also incurs $r$ (replication factor) times of network resources and may or may not be beneficial \cite{jain2005using}. 

Therefore, in this paper, instead of data replication or data redundancy, we focus on maximally improving the probability of data delivery by sending data segments through multiple paths (i.e., offloading data segments to other nodes), which leads to no additional network overhead and better performance; i.e., we choose the path(s) that has or have the maximum probability of data delivery, considering sending data fragments via multiple paths.

\subsection{Problem Statement}
Suppose node $u$ needs to transmit a data item $S$ to node $v$ within time constraint $T$. The \emph{cooperative offloading} problem is defined as to maximize the probability of successful delivery of $S$ within $T$ by offloading data among opportunistically connected nodes, where $S$ can be fragmented into different segments and each segment is transmitted over different path (i.e., each path at most carries one segment). Then, cooperative offloading can be mathematically formulated as
\begin{equation}
\label{eq:problem}
\textstyle{
\begin{split}
\max        & \quad \prod_{i=1}^{m}\prod_{j=1}^{n} P(S_i,j)^{x_{i,j}}            \\
\emph{s.t.} & \quad \sum_{i=1}^m S_i = S,\quad m \in \{1,2,\cdots,n\},          \\
            & \quad \sum_{j=1}^n x_{i,j}=1, 
            \quad \sum_{i=1}^m x_{i,j} \in \{0,1\}, 
            \quad x_{i,j} \in \{0,1\},
\end{split}
}
\end{equation}
where $n$ is the number of paths between $u$ and $v$, $m$ is the number of data segments, $P(S_i,j)$ is the probability that node $u$ can transmit segment $i$ with size $S_i$ to node $v$ within $T$ along path $j$, $x_{i,j}=1$ if segment $i$ is assigned to path $j$, otherwise $x_{i,j}=0$. Moreover, $\sum_{j=1}^n x_{i,j}=1$ ensures that each segment is assigned once, and $\sum_{i=1}^m x_{i,j} \in \{0,1\}$ ensures that each path carries at most one segment. Thus, the main problem of cooperative offloading is determining how to fragment the data item (including the number of segments and the size of each segment) and on which path to transmit each segment so as to maximize the successful delivery probability of the entire data item before its deadline. Thus, it is a joint optimization of the number of data segments, the size of each segment, and the selection of path for each segment. Although the formulation of cooperative offloading is straightforward, it turns out to be NP-hard.

\subsection{NP-hardness}
\label{sec:np}

The NP-hardness of cooperative offloading can be proven by reduction to the \emph{minimization knapsack} problem. In cooperative offloading, maximizing $\prod_{i=1}^{m}\prod_{j=1}^{n} P(S_i,j)^{x_{i,j}}$ is equivalent to minimizing
\begin{equation}
\nonumber \sum_{i=1}^{m}\sum_{j=1}^{n} -\log P(S_i,j) x_{i,j},
\end{equation}
where $-\log P(S_i,j)>0$ since $0<P(S_i,j)<1$. For a pair of segment $i$ and path $j$, it has a positive value of $-\log P(S_i,j)$ and a size $S_i$. Therefore, cooperative offloading can be rewritten as:
\begin{equation*}
\textstyle{
\begin{split}
\min        & \quad \sum_{k=1}^{mn} v_k x_k \\
\emph{s.t.} & \quad \sum_{k=1}^{mn} s_k x_k \ge S,\quad x_k \in \{0,1\},
\end{split}
}
\end{equation*}
where $v_k$ and $s_k$ respectively correspond the positive value and the size of a pair of segment and path. This is the minimization knapsack problem, which is NP-hard, and thus cooperative offloading is also NP-hard. Although cooperative offloading also admits PTAS (Polynomial Time Approximation Scheme) as the minimization knapsack problem, PTAS for cooperative offloading requires the knowledge of the value (the probability) for any pair of arbitrary data segment and path. That is, assuming the size of a data item is $S$, it needs to iteratively calculate the probability of any data segment smaller than or equal to $S$ for every path between source and destination. 
However, in opportunistic mobile networks, the estimation of such probability requires high computation as we will discuss in Section \ref{sec:framework}. Since all the computation needs to be performed on the source node (i.e., a mobile device) that has a limited computation capability, PTAS is not affordable for mobile devices and thus we design better algorithms.

\section{Probabilistic Framework}
\label{sec:framework}

In opportunistic mobile networks, even the estimation of the delivery probability of a data item along a particular path is hard. Thus, in this section, we propose a probabilistic framework to estimate such probability based on node contact pattern. The existing work \cite{jain2005using}\cite{zhuo2011contact} simplifies the estimation by considering only one node contact along an opportunistic path before the deadline and thus severely underestimates the data delivery probability. To cope with this, we consider an arbitrary number of node contacts, and thereby our framework can accurately estimate the probability.

In opportunistic mobile networks, contact patterns between nodes have been well analyzed. Similar to \cite{balasubramanian2007dtn}\cite{gao2012social}, \textit{we model the contact process between each pair of nodes as the independent Poisson process}. This modeling has been experimentally validated in \cite{gao2012social}. Moreover, unlike \cite{lin2008stochastic}\cite{lin2008efficient} that assume only a fixed amount of data can be transferred during a node contact, or \cite{yuan2009predict}\cite{gao2012social} that assume an arbitrary size of data can be sent during a node contact, we do not make such assumptions; i.e. the data amount that can be transmitted during a node contact depends on the contact duration as in \cite{jain2005using}\cite{zhuo2011contact}. In \cite{jain2005using}\cite{zhuo2011contact}, the data amount that can be transmitted between nodes is estimated based on single node contact. However, given a time constraint, nodes may contact each other multiple times and thus such an estimation is not technically sound. Therefore, to accurately estimate the delivery probability, we consider the data amount that is brought by any arbitrary number of contacts between nodes within a time constraint. Moreover, similar to \cite{zhuo2011contact}, \textit{we model contact duration between each pair of nodes as the Pareto distribution (based on per node pair statistics)}. Note that the distribution of contact frequency and contact duration is heterogeneous across different pairs of nodes; i.e., each pair of nodes has specific parameters for its distributions.

In the following, we first introduce opportunistic contact probability and data transfer probability, and then based on these, we show how to calculate data delivery probability.


\subsection{Opportunistic Contact Probability}
Since node contacts are independent Poisson processes, let random variable $T_{1}$ represent inter-contact duration between node $u$ and $v$, which follows an Exponential distribution with rate parameter $\lambda_{1}$, i.e. $T_1 \sim \mathrm{Exp}(\lambda_{1})$, where $\lambda_{1}$ is the rate parameter of the Poisson process between $u$ and $v$. Similarly, we have $T_2 \sim \mathrm{Exp}(\lambda_{2})$ for node $v$ and $w$. Then, the \textit{available probability} (denoted as $Q$) of the opportunistic path from $u$ to $w$ via $v$ (i.e., $u$ contacts $v$ first and then $v$ contacts $w$) within $T$ can be represented as $Q=P(\mathsf{T} \leq T)$, where $\mathsf{T} = T_1+T_2$. Note that the sequence of node contact is already captured by $P(\mathsf{T} \leq T)$. Assuming the probability density functions (PDF) of $T_1$ and $T_2$ are $f_1(t)$ and $f_2(t)$, respectively, $P(\mathsf{T} \leq T)$ can be calculated through the convolution $f_1(t) \otimes f_2(t)$. However, a data item may not be completely transmitted during one contact between neighboring nodes, may but require multiple contacts. Thus, multiple contacts at each hop should be considered.

Assume a data item can be transferred from $u$ to $w$ by the number of contacts $a$ between $u$ and $w$, and from $w$ to $v$ by the number of contacts $b$ between $w$ and $v$. For a collection of independent and identically distributed (\emph{i.i.d.}) random variables $\{T_i,\,\, i=1,\ldots,a\}$ and $T_i \sim \mathrm{Exp}(\lambda_{1})$, we have $\mathcal{T}_1 \sim \mathrm{Gamma}(a,\lambda_1)$, where $\mathcal{T}_1= \sum_{i=1}^a T_i$. Similarly, for $\{T_j,\, j=1,\ldots,b\}$ and $T_j \sim \mathrm{Exp}(\lambda_{2})$, we have $\mathcal{T}_2 \sim \mathrm{Gamma}(b,\lambda_2)$, where $\mathcal{T}_2= \sum_{j=1}^b T_j$. Then the PDF of $\mathbb{T}=\mathcal{T}_1+\mathcal{T}_2$ can be represented as
\begin{equation}
\label{eq:twocon}
\begin{split}
\nonumber f(t) &= f(t;a,\lambda_1) \otimes f(t;b,\lambda_2) \\
               &= \frac{t^{a-1}e^{-t\lambda_1}}{\lambda_1^{-a}\Gamma(a)} \otimes \frac{t^{b-1}e^{-t\lambda_2}}{\lambda_2^{-b}\Gamma(b)}.
\end{split}
\end{equation}

For a $k$-hop opportunistic path with the rate parameter $\lambda_i$ and the contact number $n_i$ at each hop $i$ (we also call it $k$-hop opportunistic contact path), the PDF of $\mathbb{T}=\mathcal{T}_1+\cdots+\mathcal{T}_k$ can be easily generalized as
\begin{equation}
\label{eq:ncon}
\begin{split}
f(t) &= f(t;n_1,\lambda_1) \otimes f(t;n_2,\lambda_2) \otimes \cdots \otimes f(t;n_k,\lambda_k) \\
     &= \frac{t^{n_1-1}e^{-t\lambda_1}}{\lambda_1^{-n_1}\Gamma(n_1)} \otimes \frac{t^{n_2-1}e^{-t\lambda_2}}{\lambda_2^{-n_2}\Gamma(n_2)} \otimes \cdots \otimes \frac{t^{n_k-1}e^{-t\lambda_k}}{\lambda_k^{-n_k}\Gamma(n_k)}.
\end{split}
\end{equation}
However, it is hard to calculate \eqref{eq:ncon} due to the higher order convolution. To simplify the calculation, we have the following by extending the Welch-Satterthwaite approximation to the sum of gamma random variables
\begin{equation}
\label{eq:availprob}
f(t) \approx \frac{t^{\gamma-1}e^{-t\delta}}{\delta^{-\gamma}\Gamma(\gamma)},
\end{equation}
where
\begin{equation}
\nonumber \gamma = \frac{\left(\sum_{i=1}^k n_i\lambda_i\right)^2}{\sum_{i=1}^k n_i\lambda_i^2}\,\,\, \text{ and } \,\,\,\delta= \frac{\sum_{i=1}^k n_i\lambda_i^2}{\sum_{i=1}^k n_i\lambda_i}.
\end{equation}
Thus, $\mathbb{T}$ can be approximated by a single gamma random variable $\widetilde{\mathbb{T}} \sim \mathrm{Gamma}(\gamma,\delta)$, where $\widetilde{\mathbb{T}}$ and $\mathbb{T}$ are proved to have the same mean and variance. Then, the opportunistic contact probability with time constraint $T$, denoted as $P(\mathbb{T} \leq T)$, can be easily calculated. 


It is shown in \cite{karagiannis2010power} that in some mobile traces the exponential distribution of inter-contact duration between nodes may feature a head that likely follows a power law. This dichotomy is separated by a characteristic time \cite{karagiannis2010power}, denoted as $T_c$. When the time constraint $T$ is longer than or equal to $T_c$, the probability sits on the tail of the distribution, which is exponential, and thus it is accurate. When $T$ is shorter than $T_c$, the probability sits on the head of the distribution, and thus it may be overestimated. The rigorous mathematical analysis on the mobile traces featured with the dichotomy is out of the scope of this paper.

\subsection{Data Transfer Probability} 
As the contact duration is modeled as the Pareto distribution, it can be easily concluded that the amount of data that can be transferred during a contact between a pair of nodes also follows a Pareto distribution since the data rate between a pair of nodes is relatively stable as shown in \cite{zhuo2011contact}. Let random variable $D_1$ represent the amount of data transmitted during a contact between node $u$ and $v$, which follows the Pareto distribution with shape parameter $\alpha$ and scale parameter $\beta$ ($\beta$ is the minimum possible value of $D_1$), i.e. $D_1 \sim \mathrm{Pareto(\alpha,\beta)}$. For a collection of \emph{i.i.d.} random variables $\{D_i,\,i=1,\ldots,c\}$ and $D_i \sim \mathrm{Pareto}(\alpha,\beta)$, let $\mathcal{D}=\sum_{i=1}^c D_i$. Then, the probability that a data item with size $D$ can be transferred with the number of contacts $c$ between node $u$ and $v$ is represented as $P(\mathcal{D} \geq D)$ and can be derived from the PDF of $\mathcal{D}$. However, since $\mathcal{D}$ is the sum of an arbitrary number of random variables, its PDF cannot be easily approximated by stable distributions. Therefore, we choose to approximate $\mathcal{D}$ by the maximum value of $\{D_i,\,i=1,\ldots,c\}$ denoted as $\mathcal{M}$, which relies on the observation that $\mathcal{M}$ has the same order of magnitude as the sum $\mathcal{D}$ since the Pareto distribution is a heavy-tailed distribution.

To capture such intuition, let us define $R=\mathcal{D}/\mathcal{M}$. Then, $\mathcal{D}$ can be approximated by $\mathcal{M}$, considering the ratio $R$. Then, $P(\mathcal{D} \geq D)$ can be approximated as
\begin{equation}
\label{eq:DTP}
P(\mathcal{D} \geq D) \approx 1 - \left(1 - (\frac{\beta\bar{R}}{D})^\alpha\right)^c,
\end{equation}
where $\bar{R}$ is the expectation of $R$ (see Appendix \ref{sec:appendix} for the derivation). Approximating $\mathcal{D}$ with $\mathcal{M}$ bears a low relative error about 10\% \cite{Zaliapin2005approx}.

\subsection{Data Delivery Probability} 
Given an opportunistic path, a data item $D$ and a time constraint $T$, we will show how to calculate the probability of successful delivery of $D$ over the opportunistic path within $T$, denoted as $P(T,D)$, based on the opportunistic contact probability and data transfer probability. 

First, we consider the case of a one-hop path. The data item needs at most $l= \lceil \frac{D}{\beta} \rceil $ node contacts to be completely transferred. Then, $P(T,D)$ can be calculated as the sum of the probabilities that the data item is completely transmitted at each particular contact as following
\begin{equation}
\label{eq:onehop}
P(T, D) = \sum_{i=1}^l \widetilde{P}_{i-1} \cdot P(\mathbb{T}_i \leq T - T^{'}) \cdot P(\mathcal{D}_i \geq D),
\end{equation}
where
\begin{eqnarray}
\nonumber\widetilde{P}_i =
\left\{\begin{matrix}
  \underset{j=1}{\overset{i}{\prod}} P(\mathbb{T}_j \leq T - T^{'}) \cdot P(\mathcal{D}_j < D)   & i>0
  \\
 1 & i=0,
\end{matrix}\right.
\end{eqnarray}
$T^{'}=\frac{D}{r}$ denotes the transmission time of the data item and $r$ is the data rate between these two nodes. As in \eqref{eq:onehop}, the probability that $i$th contact completes the data transfer can be interpreted as the probability that the data can be transferred within $i$ node contacts when  $i-1$ contact fails (i.e. the data transferred by former $i-1$ contacts is less than $D$ and the sum of inter-contact duration of $i$ contacts is less or equal to $T-T^{'}$). 

Then, we can extend the calculation of $P(T,D)$ to a $k$-hop path. Given a $k$-tuple
\begin{equation}
\label{eq:tuple}
\langle n_1, \ldots,n_i,\ldots,n_k \rangle, \,\, 1 \leq n_i \leq \lceil \frac{D}{\beta_i} \rceil,
\end{equation}
which falls into one of the $\prod_{i=1}^k \lceil \frac{D}{\beta_i} \rceil$ possible combinations, we need to calculate the probability that the data can be transferred through the path by the specified number of contacts at each hop as $n_i$ in the $k$-tuple. Similar to the calculation for the one-hop path, we need to consider the failure probability of tuples which have one less contact number at one of the hops. For example, for $2$-tuple $\langle n_1, n_2 \rangle$, we need consider the failure probability of both $\langle n_1-1,n_2 \rangle$ and $\langle n_1, n_2-1 \rangle$. Finally, $P(T,D)$ on the $k$-hop path can be calculated as the sum of the probabilities of all the combinations, and the computational complexity is $\prod_{i=1}^k \lceil \frac{D}{\beta_i} \rceil$. $\beta$ depends on the contact duration and data transmission rate between nodes, and thus it varies in different application scenarios. The complexity depends on not only $\beta$ but also $D$. Given a network, $\beta$ between node pairs is in the same scale based our analysis on real traces (e.g., in MIT Reality trace \cite{eagle2006reality}, the minimum contact duration of 99\% node pairs is between 5 and 50 minutes) and hence $D$ determines the complexity. Moreover, $D$ should not be much larger than $\beta$. Given a typical value of $\alpha$, e.g., 2 as in MIT Reality trace, it approximately needs $D/2\beta$ contacts to transfer $D$ between a pair of nodes. Assuming $D$ and $\beta$ in different scales, e.g., $D=10$GB and $\beta=1$MB. We do not expect any applications can tolerate a data transfer to be completed by 5000 opportunistic contacts. Therefore, the computational complexity is low in practice. 

Note that we do not consider the case where an edge can be shared by multiple paths, since this will complicate the problem and make it impossible to calculate the data delivery probability. 

In summary, the probabilistic framework explores contact pattern between nodes to provide the estimation of data delivery probability over the opportunistic path, which is the basis of our designed algorithms. To the best of our knowledge, this is the first work that provides such estimation without any restrictions.

\section{Heuristic Algorithm}
\label{sec:Heuristic}
In this section, based on the proposed probabilistic framework, we give the design of the heuristic algorithm. 

\subsection{The Algorithm}
According to (\ref{eq:problem}), intuitively, a good solution of cooperative offloading should have a small number of data segments while keeping a high delivery probability for each segment. Therefore, the basic idea of the heuristic algorithm is to first find some paths between source and destination that have a higher available probability than the direct path. Then, it is determined how much data should be assigned at each of these paths, while maintaining a high data delivery probability. Finally, data reallocation among these paths is performed to maximally improve the delivery probability of the entire data item.

To choose particular paths for data offloading, we need to measure the capability of paths for data transmission. Since all paths are opportunistic, which are characterized by contact frequency and contact duration at each hop, we can explore these two properties to measure the capability of each path. However, without prior knowledge of how much data will be assigned at each path, it is difficult to quantify the capability of paths. Therefore, we employ the available probability of the opportunistic path within the time constraint $Q$ as the metric to characterize each path, since the establishment of the opportunistic path within the time constraint is the prerequisite of data transmission between the pair of nodes. Moreover, the available probability of direct path (one hop path between source and destination), denoted as $Q^{'}$, is used as the criterion to choose these paths.

After allocating the paths, we need to determine how much data should be assigned to each path initially. When nodes contact each other, $\beta$ is the size of data that can be guaranteed to be transferred (according to the Pareto distribution). Thus, for each path, we allocate the amount of data that can be guaranteed to be transferred along the path if the path can be established with the time constraint, i.e. $\min\{\beta_i,\, i=1,\cdots,k \}$. We call this the \emph{path capacity}, denoted as $C_{uv}^i$ for path $i$ between node $u$ and $v$. Note that the delivery probability of initially assigned data at a selected path is the same with the available probability $Q$ of the path, so no additional calculation is needed.

\begin{algorithm}[!b]
\begin{footnotesize}
\caption{Heuristic Algorithm}
\label{alg:Probability}
\SetKwFunction{DijkstraMaxCapacity}{DijkstraMaxCapacity}
\SetKwFunction{AllocateResource}{ExcludeAllocPath}
\SetKwFunction{DijkstraMaxQ}{DijkstraMaxQ}
\SetKwFunction{ReleaseResource}{ReleaseResource}
\SetKwFunction{CalculateProb}{CalculateProb}
\SetKwFunction{Selection}{Selection}
\SetKwInOut{Input}{Input}
\SetKwInOut{Output}{Output}
\SetKwRepeat{DoWhile}{do}{while}

\Input{$u,v,S,T$}
\Output{$\mathcal{P}_A$, $\mathcal{S}$}

\While{\DijkstraMaxQ{$u,v$} $\neq$ $\emptyset$ \&\& $\sum_{i \in \mathcal{P}_A} S_i < S$}
{
	$p \leftarrow$ \DijkstraMaxQ{$u,v$}  \\
	\If{$Q_p < Q^{'}$}{
		\textbf{break} \\
	}
	\AllocateResource{$p$}                   \\
	$\mathcal{P}_A \leftarrow \mathcal{P}_A \cup \{p\}$ \\
	$\mathcal{S} \leftarrow \mathcal{S} \cup \{S_p\}$  \\
}

\While(\tcp*[h]{\scriptsize assign remaining data}){$S-\sum_{i \in \mathcal{P}_A} S_i >0$}
{
	$p \leftarrow \arg\max_{i \in \mathcal{P}_A} P_{uv}^i(T,S_i)$ \\
	$q \leftarrow \arg\max_{i \in \mathcal{P}_A \backslash\{p\}} P_{uv}^i(T,S_i)$ \\
	\While{$P_{uv}^p(T,S_p) \geq P_{uv}^q(T,S_q)$ \&\& $S-\sum_{i \in \mathcal{P}_A} S_i >0$}
	{
		$S_p \leftarrow S_p + \Delta$ \tcp*[h]{\scriptsize $\Delta$ is increment of  assigned data}
	}
}
\While(\tcp*[h]{\scriptsize reallocate data}){}
{
	$\mathcal{P}_A^{'} \leftarrow \mathcal{P}_A$, $\mathcal{S}^{'} \leftarrow \mathcal{S}$ \\
	$j \leftarrow \arg\min_{i \in \mathcal{P}_A^{'}} P_{uv}^i(T,S_i)$ \\
	$\mathcal{P}_A^{'} \leftarrow \mathcal{P}_A^{'} \backslash \{j\}$, $\mathcal{S}^{'} \leftarrow \mathcal{S}^{'} \backslash \{S_j\}$ \\
	\While{$S_j >0$}
	{
		$p \leftarrow \arg\max_{i \in \mathcal{P}_A^{'}} P_{uv}^i(T,S_i)$ \\
		$q \leftarrow \arg\max_{i \in \mathcal{P}_A^{'} \backslash\{p\}} P_{uv}^i(T,S_i)$ \\
		\While{$P_{uv}^p(T,S_p) \geq P_{uv}^q(T, S_p)$ \&\& $S_j >0$}
		{
			$S_p \leftarrow S_p + \Delta$ \\
		    $S_j \leftarrow S_j - \Delta$ \\
		}
	}
	\eIf{$\prod_{i \in \mathcal{P}_A} P_{uv}^i(T,S_i) < \prod_{i \in \mathcal{P}_A^{'}} P_{uv}^i(T,S_i)$}
	{
		$\mathcal{P}_A \leftarrow \mathcal{P}_A^{'}$, $\mathcal{S} \leftarrow \mathcal{S}^{'}$ \\
	}
	{
		\textbf{break} \\
	}
	
}
\end{footnotesize}
\end{algorithm}

Considering an example where node $u$ needs to transmit a data item $S$ to node $v$ within $T$, the heuristic algorithm works as follows. First, we adopt Dijkstra's algorithm with the metric of $Q$ to find the path from node $u$ to $v$ that has the minimum $1/Q$. Note that in Dijkstra's algorithm $Q$ is calculated for the path between node $u$ and unvisited node according to \eqref{eq:availprob}. If the found path $i$ has $Q_i \geq Q^{'}$ (if node $u$ does not have a direct path to $v$, $Q^{'}=0$), then we assign the data amount of $C_{uv}^i$ to path $i$, i.e., $S_i=C_{uv}^i$ if $C_{uv}^i < S - \sum_{i \in \mathcal{P}_A} C_{uv}^i$, otherwise $S_i=S-\sum_{i \in \mathcal{P}_A} C_{uv}^i$, and add $i$ into the set of allocated paths $\mathcal{P}_A$. After allocating path $i$, the edges along the path will be removed from the network and not considered in subsequent searches. 

The searching process is iteratively executed until it meets one of the following stop conditions: \textit{(i)} $Q_i < Q^{'}$; \textit{(ii)} node $v$ cannot be reached by Dijkstra's algorithm from $u$; \textit{(iii)} $\sum_{i \in \mathcal{P}_A} S_i = S$. If the searching process ends with $|\mathcal{P}_A|=0$, there is no need to offload and it is better to send the data item directly from $u$ to $v$. If $\sum_{i \in \mathcal{P}_A} S_i < S$, the remaining data needs to be assigned to $\mathcal{P}_A$. If $\sum_{i \in \mathcal{P}_A} S_i = S$, data reallocation among $\mathcal{P}_A$ is needed to improve the delivery probability of $S$. The path searching process is from line 1 to 9 in Algorithm \ref{alg:Probability}

To assign the remaining data of $S-\sum_{i \in \mathcal{P}_A} C_{uv}^i$ to $\mathcal{P}_A$, first we rank $\mathcal{P}_A$ according to the delivery probability of the assigned data at each path, i.e. $P_{uv}^i(T,S_i),\,i \in \mathcal{P}_A$. Then, we assign more data to the path with the highest delivery probability among $\mathcal{P}_A$ until the probability is lower than the second highest one. The process is repeated until there is no remaining data. When assigning more data to the selected allocated path, we need to determine the amount of data to be assigned each time. As the path capacity of a $k$-hop path is $\min\{\beta_1,\beta_2,\ldots, \beta_k\}$, each time we increase the amount of the assigned data to next higher value among $\{\beta_1,\beta_2,\ldots, \beta_k\}$. If the assigned data is more or equal to $\max\{\beta_1,\beta_2,\ldots, \beta_k\}$, each time the assigned data is increased by the path capacity. The assignment of the remaining data is from line 10 to 16 in Algorithm \ref{alg:Probability} 

After all the remaining data is assigned, we further refine the delivery probability of $S$ by reallocating the data assigned at the path, say path $i$, which has the lowest delivery probability among $\mathcal{P}_A$, to other paths in $\mathcal{P}_A$ using the same approach described above.
If the reallocation of $S_i$ improves the delivery probability of $S$, path $i$ is excluded from $\mathcal{P}_A$ and the process is repeated, otherwise the reallocation process stops (from line 29 to 33 in Algorithm \ref{alg:Probability}). The heuristic algorithm ends up with the allocated paths and the data assignment at each path .

\subsection{Discussion}
The heuristic algorithm runs on mobile devices. When each mobile device connects to the infrastructure, it uploads the contact information with other nodes to the infrastructure such that the infrastructure has the global information of the network. Then, the infrastructure sends out the up-to-date global information each time when a node is connected to it such that each node has the global information. When a node needs to transmit data to the infrastructure, it can run the heuristic algorithm based on the global information it currently has to determine how to cooperatively offload the data. The global information may reveal users' contacts and intrude on their privacy. However, user's identity is hidden behind node \textit{id}. Without correlating a user with node \textit{id}, user's privacy is secure. More sophisticated and rigorous treatment of privacy control is out of the scope of this paper.

The contact information between a node pair is represented by three parameters, i.e., $\alpha$, $\beta$, and $\lambda$. Let $d_m$ denote the maximum node degree and $N$ be the number of nodes in the network. The size of exchanged contact information during a contact between a node and infrastructure is at most $3d_mN$. Since the maximum node degree $d_m$ commonly does not increase with the network size $N$, the overhead of exchanged contact information linearly scales with the network size. 

In the algorithm, we exclude the allocated paths from the subsequent path searches, because it is extremely hard to calculate the data delivery probability if multiple data segments are transmitted along an edge which is shared by multiple allocated paths. Therefore, we choose to avoid the difficulty by regulating an edge to be occupied by only one allocated path. Moreover, this also limits the maximum number of allocated paths to the minimum node degree (number of neighbors) of source and destination, and thus path allocation using Dijkstra's algorithm can terminate quickly. 

Intuitively, multi-hop paths should be considered for data transmission when its available probability is higher than a one-hop path. Therefore, $Q'$ is employed as a criterion to allocate paths. As indicated in Section \ref{sec:framework}, the computational complexity of data delivery probability over multi-hop paths increases with the number of hops. However, due to the adoption of $Q'$, allocated paths are usually limited to paths with fewer hops and thus we can avoid the high computational complexity of delivery probability when data that is more than the path capacity is assigned to a multi-hop path.

Data assignment at each allocated path is initially set to the path capacity, and thus the data delivery probability at each path is the same as $Q$. This is a good choice because we do not know which paths are sensitive (in terms of delivery probability) to the increase of assigned data beforehand. After path allocation and initial data assignment, the number of data segments is the same as that of allocated paths. Then, data is reallocated from paths of lower delivery probability to ones of higher probability so as to improve the delivery probability of the data item by reducing the number of allocated paths. This complies with the design intuitions that the number of data segments should be small and each allocated path should have a high delivery probability. 

Moreover, the computational complexity of the heuristic algorithm is much lower than PTAS for cooperative offloading. PTAS needs to search all possible paths (they can be very long) between source and destination; the heuristic algorithm only needs to locate the paths, the available probability of which is more than $Q'$. In addition, PTAS has to calculate the delivery probability of each path when carrying all possible amounts of data. This incurs the most of the computation. Unlike PTAS, the heuristic algorithm only needs to calculate certain amounts of data for the few allocated paths. Due to the uncertainty of the  number of paths, path length and path capacity, it is hard to give the mathematical comparison of the computational complexity between the heuristic algorithm and PTAS. However, based on the reasoning stated above, it is easy to conclude that the computation of the heuristic algorithm is much less that of PTAS, and its performance is close to the optimum as we will demonstrate in Section \ref{sec:Evaluation}.

For the heuristic algorithm itself, excluding the complexity of calculating the data delivery probability, the complexity includes three parts. (\textit{i}) The complexity of allocating paths between a pair of source and destination is $d_mN^2$. Since a selected path is excluded from the network, at most $d_m$ paths are allocated. Dijkstra's algorithm costs $N^2$ each run and thus totally $d_mN^2$. (\textit{ii}) The complexity of assigning remaining data to selected paths is $S^2$, where $S$ is the size of the data item, since the remaining data is at most $S$ and the increment of data assignment is at least 1. (\textit{iii}) The complexity of reallocating data among selected paths is $d_m S^2$, because there are at most $d_m$ reallocation processes and each costs at most $S^2$. So, the combined complexity is $d_mN^2 + S^2 + d_mS^2$. Since the maximum node degree $d_m$ commonly does not increase with the network size $N$, the complexity of the heuristic algorithm is $O(N^2+S^2)$. 

Based on the characteristics of the heuristic algorithm, we expect the use cases of the heuristic algorithm, in general, are the scenarios where contact pattern between nodes is relatively stable and communication overhead is small, which means the time taken to  exchange contact information between infrastructure and node is relatively short compared to their contact duration by jointly considering bandwidth and network size. If contact pattern between nodes varies largely, the information collected by the infrastructure may be stale, which may affect the performance. If the network size is too large, the communication overhead may dominate and little data can be transmitted from a node to the infrastructure during a contact. An example use case is disaster recovery, where rescue crews and survivors can better communicate with the command center using their smartphones via limited mobile cellular tower \cite{lu2016networking} by cooperative data offloading. The global information can be stored at the command center (directly connected with mobile cellular towers), which can be trusted in this scenario. Moreover, the heuristic algorithm is ready and easy to deploy on the delay-tolerant network architecture DTN2 \cite{dtn2} and the smartphone-based system \cite{lu2016networking} for disaster recovery.   
 
\section{Distributed Algorithm}
\label{sec:Distributed}
Global information is required for the heuristic algorithm presented above. However, in some scenarios such information might not be available or cost too much to collect. Furthermore, the pattern of contact frequency and contact duration between nodes may vary largely over time and thus the collected information may be stale, which will impact the performance of the heuristic algorithm. Thus, in this section we propose a distributed algorithm to address these problems.

In opportunistic mobile networks, it is hard to maintain multi-hop information locally, since nodes only intermittently contact each other and thus the information cannot be promptly updated. Moreover, from the probabilistic framework, we can infer that shorter paths are prone to have a higher data delivery probability, and thus maintaining multi-hop information can also be wasteful. Therefore, in the design of the distributed algorithm, it is only required that each node maintains two-hop information of contact frequency and contact duration between nodes; i.e., each node maintains contact frequency and contact duration with its neighbors, and when two nodes encounter, they will exchange such information and then they will have the two-hop information. The goal of the distributed algorithm is to determine whether and how much data should be transmitted when the node carrying the data encounters other nodes based on the collected information so as to improve the delivery probability. 

\begin{figure}[!t]
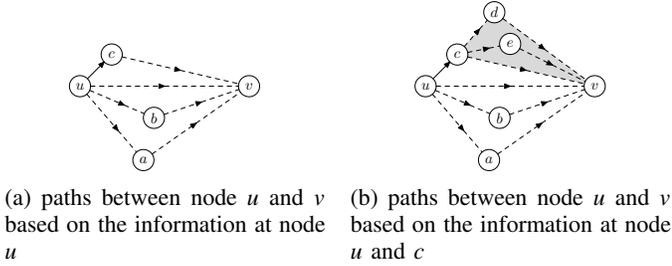

\centering
	\begin{subfigure}[t]{0.235\textwidth}
		\centering
    	\includegraphics[width=.6\textwidth]{MobileOffloading.9}
    	\caption{paths between node $u$ and $v$ based on the information at node $u$}
    	\label{fig:PathAtS}
    \end{subfigure}
    \hspace{0.1cm}
    \begin{subfigure}[t]{0.235\textwidth}
    	\centering
    	\includegraphics[width=.6\textwidth]{MobileOffloading.13}
    	\caption{paths between node $u$ and $v$ based on the information at node $u$ and $c$}
    	\label{fig:PathAtSandC}
    \end{subfigure}
\caption{Paths constructed based on local information}
\label{fig:DistributedPaths}
\end{figure}

With the two-hop information maintained locally, the source node can initially construct paths to the destination (infrastructure). For example, as shown in Figure \ref{fig:PathAtS}, there are four paths that can be constructed from node $u$ to node $v$ including one one-hop path and three two-hop paths, where node $a$, $b$ and $c$ are the potential nodes to cooperate for data transmission. Moreover, when the source node encounters a neighbor, it also has the local information maintained at this neighbor. For example, as shown in Figure \ref{fig:PathAtSandC}, when node $u$ encounters node $c$, node $u$ learns the information of the two-hop paths between $c$ and $v$, and then node $u$ can construct more paths to node $v$.

The distributed algorithm exploits the locally maintained information for data offloading, which includes three phases: \textit{criterion assignment}, \textit{real-time adjustment} and \textit{assignment update}. The first phase leverages the local information to select the paths for offloading and determines the data assignment for each path; the second phase adjusts the data assignment based on the local information of the encountered node to achieve a better delivery probability; the third phase updates the data assignment at the sender and receiver based on the actually transmitted data amount between them. Consider Figure \ref{fig:DistributedPaths} as an example, where node $u$ has data $S$ to transmit to $v$. First, node $u$ determines the criterion assignment based on the paths in Figure \ref{fig:PathAtS}. Then, when node $u$ encounters $c$, node $u$ needs to adjust the size of data to be transmitted to node $c$, which can maximally improve the delivery probability of $S$ determined by the criterion assignment, based on the additional paths in Figure \ref{fig:PathAtSandC}. After the data transmission between $u$ and $v$, both need to update their data assignment.

\subsection{Criterion Assignment}
Although the data to be transmitted to a neighbor will be adjusted when the source node meets the neighbor, the source node should provide a reasonable criterion for later adjustment. To obtain a good delivery probability, the criterion assignment is determined as follows. Let $\mathcal{P}_{uv}$ denote the set of paths from node $u$ to $v$, which only includes the paths that are no more than two hops from node $u$ to $v$, and let $\mathcal{C}_{uv}$ denote the capacity of $\mathcal{P}_{uv}$, where $\mathcal{C}_{uv}=\sum_{i \in \mathcal{P}_{uv}} C_{uv}^i$. If $\mathcal{C}_{uv} < S$, the data assignment at path $i$ in $\mathcal{P}_{uv}$ is $S_i = S \times \frac{C_{uv}^i}{\mathcal{C}_{uv}}$. If $\mathcal{C}_{uv} \geq S$, we rank $\mathcal{P}_{uv}$ according to the available probability of each path, and then the first $k$ paths of $\mathcal{P}_{uv}$, where $\sum_{i=1}^k C_{uv}^i \geq S$, are selected and the assigned data for each path is equal to the path capacity except the last one, the assignment of which is $S - \sum_{i=1}^{k-1} C_{uv}^i$. Let $\mathcal{A}_u$ denote the data assignment at node $u$, $\mathcal{A}_u=\{S_i,\,i \in \mathcal{P}_{uv}\}$. This data assignment is determined before transmission. 

\subsection{Real-time Adjustment}
With the criterion assignment, we have the data assignment for each path from node $u$ to $v$. The data assigned at the path that goes through node $c$ is the amount of data to be transmitted to node $c$ when nodes $u$ and $c$ encounter. However, node $c$ may be more likely to deliver a certain amount of data to node $v$ than node $u$. If so, this amount of data should be transmitted to node $v$ from $c$ instead of $u$ so as to improve the delivery probability. Thus, node $u$ needs to investigate this capability of node $c$ based on $c$'s local information (paths to node $v$ as the shadow part in Figure \ref{fig:PathAtSandC}) when they are in contact, and then adjust the amount of data to be transmitted to $c$ accordingly.

The real-time adjustment works as follows. When nodes $u$ and $c$ are in contact, first $u$ selects the path among $\mathcal{P}_{uv}$ that has the minimum delivery probability of the assigned data before deadline. Let us say the selected path is $j$ and its assigned data is $S_j$. Then, the path among $\mathcal{P}_{cv}$, to which the reallocation of $S_j$ can maximally improve the delivery probability of $S$, can be determined, say path $k$. $S_j$ should be transferred to node $v$ from $c$ instead of $u$, and thus $S_j$ needs to be transmitted from $u$ to $c$ first. So the data amount to be transmitted from $u$ to $c$ will be increased by $S_j$ and $S_j$ will be assigned at path $k$. The process is repeated. If the reallocation of $S_j$ cannot increase the delivery probability of $S$, the process stops.

The following shows how to compare the improvement of the delivery probability of $S$ when $S_j$ is reassigned to different paths, say path $d$ and $e$ in $\mathcal{P}_{cv}$. Let $S_d$ and $S_e$ denote the amount of data already assigned at path $d$ and path $e$, respectively, which can be the assignment of data currently carried by node $c$ or zero. To compare their improvement, we only need to calculate the delivery probabilities of $S_j+S_d+S_e$ when $S_j$ is reallocated to path $d$ and path $e$, respectively, since the assignment of the rest of $S$ is not impacted. If $S_j$ is reassigned at path $d$, the delivery probability of $S_j+S_d+S_e$ is the product of the delivery probability of $S_d+S_j$ along path $d$ and the delivery probability of $S_e$ along path $e$. Similarly we can calculate the delivery probability of $S_j+S_d+S_e$ if $S_j$ is assigned at path $e$, and then we can determine to which path $S_j$ should be assigned.

\subsection{Assignment Update}
The real-time adjustment ends with the total amount of data to be transmitted from node $u$ to node $c$, denoted as $S_c$, and the assignment of $S_c$ at $\mathcal{P}_{cv}$. Together with the assignment of currently carried data at node $c$, let $\mathcal{A}_c$ denote the total  data assignment at node $c$, $\mathcal{A}_c=\{S_i, i \in \mathcal{P}_{cv}\}$. As the actually transmitted data from $u$ to $c$, denoted by $S_c^{'}$, may be less than $S_c$ due to the uncertainty of their contact duration, both node $u$ and node $c$ need to update the data assignment after the data transmission. For node $u$, the amount $S_c^{'}$ should be excluded from the assignment $\mathcal{A}_u$ by sequentially removing the data assigned at the path with the lowest delivery probability. Similarly, for node $c$, the amount $S_c-S_c^{'}$ needs to be removed from $\mathcal{A}_c$.

\subsection{Discussion}
Based on these three phases, the distributed algorithm works as follows. First, the source node determines a criterion assignment for the data item. Then, when the node that carries the data encounters a neighbor, it first determines whether the neighbor is the source node or if it received the data from this neighbor before. If so, no data will be transmitted. If the encountered node does not have the data, the node carrying the data needs to determine how much data is to be transmitted to the encountered node (if both of nodes have the data, they also need to determine who should transmit data to another) by comparing the improvement of the delivery probability by data reallocation. After that, the determined reallocated data is transmitted and then the data assignment at sender and receiver is updated based on the amount of data actually transferred during their contact.

In the design of the distributed algorithm, nodes are only required to maintain two-hop information. One motivation for this is that, as stated above, nodes opportunistically contact each other and hence locally maintained multi-hop information cannot be promptly updated. Another reason is that by regulating to two-hop paths, the data delivery probability over the paths can be easily calculated and thus nodes can quickly make decisions at runtime, i.e., at the phase of real-time adjustment.

Although the distributed algorithm requires that the source node must have a path with no more than two hops to infrastructure, it can easily establish these paths through its neighbors since most nodes in the network intermittently connect with infrastructure. Moreover, intermediate nodes, e.g. node $c$ in Figure \ref{fig:PathAtSandC}, will also exploit two-hop paths to forward data. Therefore, there are enough paths to be explored for data transmission.  

The design of the distributed algorithm takes advantage of both node contact patterns and stochastic node contacts. Criterion assignment is determined based on two-hop contact patterns, which can be interpreted as how much data is \textit{expected} to be transmitted along each path. Real-time adjustment exploits stochastic node contacts to find opportunities (e.g., additional paths) that can improve the delivery probability at runtime. These schemes make up for the lack of global information, and thus the distributed algorithm should perform close to the heuristic algorithm. 

The distributed algorithm does not require the central entity to store the global information, thus it can be widely deployed to opportunistic mobile networks. An example use case is vehicular ad hoc networks. Since the contact duration between vehicle and roadside unit is short, vehicles can leverage cooperative data offload to overcome the incomplete data transfer due to the transient contact. Since the heuristic algorithm incurs additional communication overhead between vehicle and roadside unit, it does not work well in this scenario. However, the distributed algorithm, which is insensitive to network size and has no communication overhead between vehicle and roadside unit, perfectly fits this scenario. Moreover, the distributed algorithm is ready and easy to be integrated with the architecture DTN2, where infrastructure can be treated just as a type of nodes. 

\section{Performance Evaluation}
\label{sec:Evaluation}

In this section, we evaluate the performance of the heuristic algorithm and distributed algorithm based on synthetic networks and real traces.

\begin{scriptsize}
\begin{table}[t]
\renewcommand{\arraystretch}{1.2}
\caption{Parameter settings for benchmark}
\centering
\label{tab:Settings}
\begin{tabular}{c|c|c}
        \hline
        \textbf{Parameter}  & \textbf{Value}  &  \textbf{Meaning} \\      \hline \hline
        $n$                 & 100             &  number of nodes  \\
        $\mu_w$             & 0.3			  &  mixing parameter for weights \\
        $\mu_t$				& 0.3			  &  mixing parameter for topology \\ 
        $\xi$         		& 2               &  exponent for weight distribution     \\
        $d$         		& 5, 10           &  average node degree         \\
        $d_{m}$   			& 8, 15           &  maximum node degree           \\     \hline
\end{tabular}
\end{table}
\end{scriptsize}

\begin{scriptsize}
\begin{table}[t]
\renewcommand{\arraystretch}{1.2}
\centering
\caption{Parameter settings for node contacts}
\begin{tabular}{c|c|c}
    \hline
 \textbf{Type}& \textbf{Parameter} &\textbf{Value} \\
 \hline\hline
\multirow{3}{*}{Between nodes} & $\alpha$ & $[3, 4]$, $[6, 10]$ \\
 & $\beta$ & $[2,3]$ \\
 & $\lambda$ & $1/\mathrm{edge \,weight}$ \\
\hline
\multirow{3}{*}{\begin{tabular}[c]{@{}c@{}} Between node \\ and infrastructure\end{tabular}} & $\alpha'$ & $[3,4]$ \\
 & $\beta'$ & $[2,3]$ \\
 & $\lambda'$ & $[0.001,0.1]$, $[0.01,0.1]$ \\
\hline
\end{tabular}
\label{tab:ContactSettings}
\end{table}
\end{scriptsize}

\subsection{Evaluation on Synthetic Networks}
First we investigate the performance of the heuristic algorithm on synthetic networks. The synthetic networks are generated by the well-know benchmark \cite{lancichinetti2009benchmarks}. It provides power-law distribution of node degree and edge weight, and various topology control. There are several parameters to control the generated network: the number of nodes, $n$; the mixing parameter for the weights, $\mu_w$; the mixing parameter for the topology, $\mu_t$; the exponent for the weight distribution, $\xi$; the average node degree, $d$; and the maximum node degree. $d_{m}$. The settings of these parameters are shown in Table \ref{tab:Settings}.

With the synthetic networks, we also need to generate the contact pattern between nodes and between node and infrastructure. The settings of these distribution parameters (i.e., $\alpha,\,\beta$ and $\lambda$) are shown in Table \ref{tab:ContactSettings}, where $\alpha=[3,\,4]$, for example, means that $\alpha$ between pair of nodes is set to a random number between 3 and 4, which follows the uniform distribution.  The parameter settings are chosen based on our analysis on real traces (i.e., MIT Reality and DieselNet).

When a node transmits data to the infrastructure, it needs to decide whether to offload the data to other nodes so as to improve the delivery probability. So, the node goes through the following procedure. First, it calculates the probability that the node directly transmits the data to the infrastructure (no offloading, the data is transmitted only when it connects with infrastructure), denoted as \emph{Individual Est}. Then it employs the heuristic algorithm to calculate the probability if it offloads the data to other nodes, denoted as \emph{Cooperative Est}. If \emph{Cooperative Est} is greater than \emph{Individual Est}, the node offloads the data to the selected paths with the corresponding data assignment determined by the heuristic algorithm. \emph{Individual Est} and \emph{Cooperative Est} are compared for data transmissions with different data sizes and deadlines at each node in the networks. We also compare the delivery probabilities based on simulations (denoted as \emph{Individual Sim} and \emph{Cooperative Sim}). Specifically, we generate the random numbers for inter-contact duration and contact duration between neighboring nodes according to their distributions, then data is transmitted between nodes based on these generated contact information.

\begin{figure}[!t]
\centering
	\begin{subfigure}[t]{0.235\textwidth}
		\centering
    	\includegraphics[width=.9\textwidth]{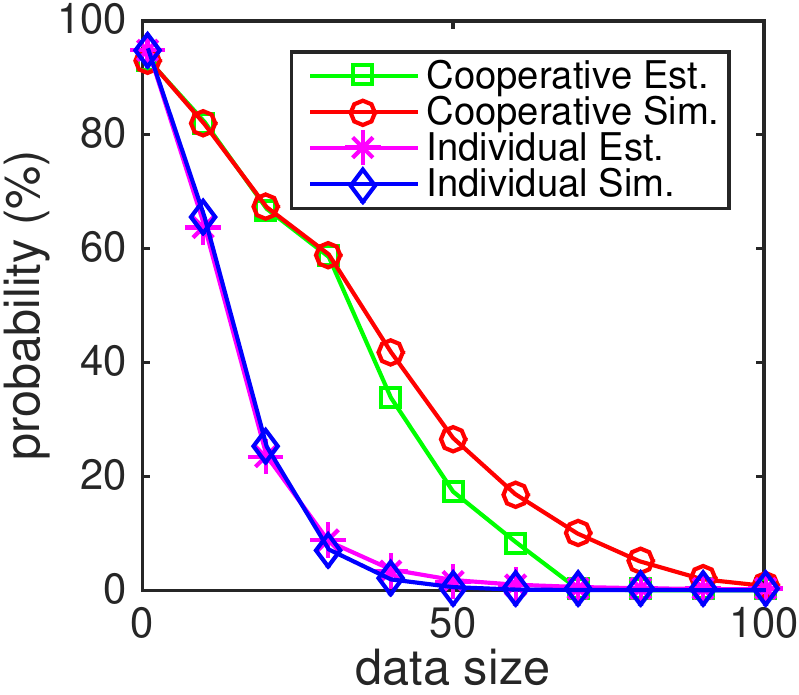}
    	\caption{probability vs data size, $T=400$}
    	\label{fig:datasize}
    \end{subfigure}
    \hspace{0.1cm}
    \begin{subfigure}[t]{0.235\textwidth}
    	\centering
    	\includegraphics[width=.9\textwidth]{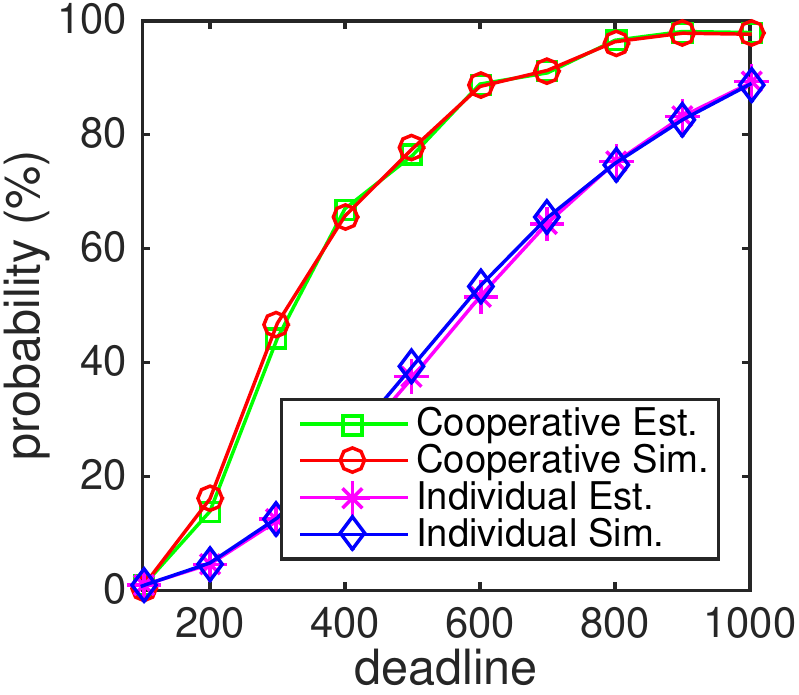}
    	\caption{probability vs deadline, $S=20$}
    	\label{fig:deadline}
    \end{subfigure}
\caption{Data Delivery probability based on estimation and simulation for \emph{Individual} and \emph{Cooperative} in synthetic networks, where $d=10$, $d_m=15$, $\alpha=[6,10]$, $\lambda'=[0.001,0.01]$.}
\label{fig:TransProb}
\end{figure}

\begin{figure}[!t]
\centering
	\begin{subfigure}[t]{0.235\textwidth}
		\centering
    	\includegraphics[width=0.9\textwidth]{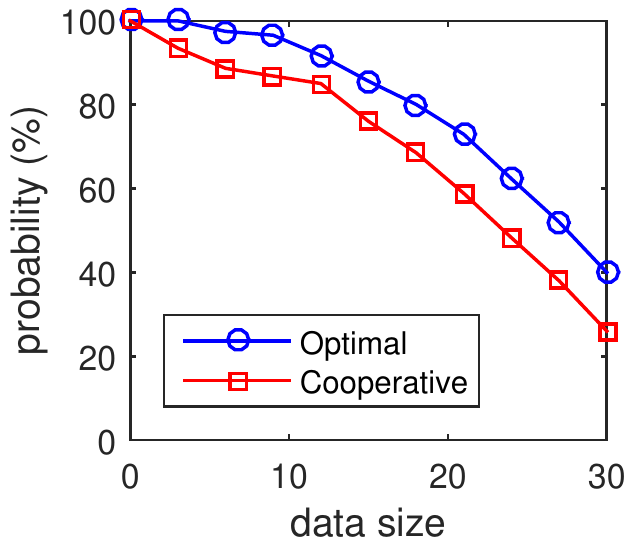}
    	\caption{probability vs data size, $T=400$}
    	\label{fig:datasizeopt}
    \end{subfigure}
    \hspace{0.1cm}
    \begin{subfigure}[t]{0.235\textwidth}
    	\centering
    	\includegraphics[width=0.9\textwidth]{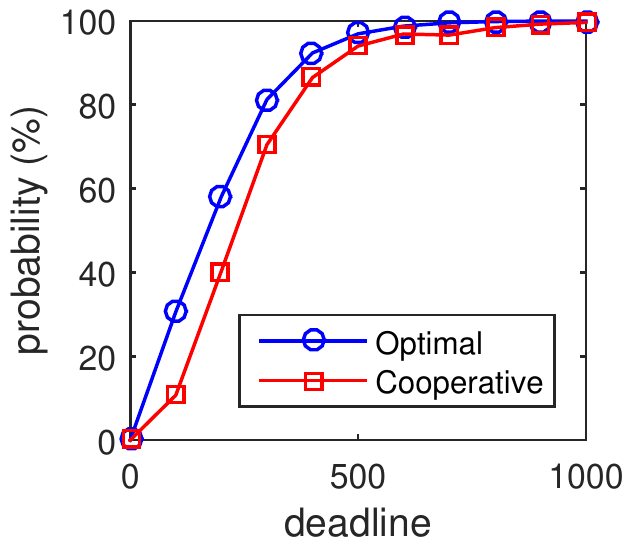}
    	\caption{probability vs deadline, $S=10$}
    	\label{fig:deadlineopt}
    \end{subfigure}
\caption{Data Delivery probability based on estimation for \emph{Optimal} and \emph{Cooperative}. Data is sent to infrastructure from a selected node in a synthetic network, where $d=10$, $d_m=15$, $\alpha=[6,10]$, $\lambda'=[0.001,0.01]$.}
\label{fig:opt}
\end{figure}

\textbf{\textit{Cooperative} does help.} Figure \ref{fig:TransProb} shows the comparison of the delivery probabilities for \emph{Individual} and \emph{Cooperative} in synthetic networks. Figure \ref{fig:datasize} shows the successful probability of transmissions with varying data sizes, meanwhile Figure \ref{fig:deadline} shows the successful probability of transmissions with varying deadlines, where the estimated probability is averaged for all the nodes for each transmission, and the simulated probability is averaged for all the nodes with 500 simulation runs (i.e., the simulated probability is the number of successful transmissions divided by the total number of transmissions). 

As shown in Figure \ref{fig:datasize}, for both \emph{Cooperative} and \emph{Individual}, the probability decreases with the increase of data size as expected. When the size of data is small, the data can be easily transmitted directly to infrastructure, and thus their probabilities are similar. However, when the size of data increases, the probability of \emph{Individual} drops dramatically from $S=10$ to $S=40$ while cooperative offloading significantly improves the delivery probability (e.g., it can be increased from 20\% to 70\% when $S=20$).  When the data size increases further, the probability of \emph{Cooperative} also decreases, i.e. cooperative offloading cannot improve the delivery probability as much as before. 
Figure \ref{fig:deadline} shows the probability of data transmissions with varying deadlines. Similarly, \emph{Cooperative} and \emph{Individual} start at the same probability. The difference between them expands and then narrows when the deadline further looses. 

\begin{figure}[!t]
\centering
    \begin{subfigure}[t]{0.48\textwidth}
    	\centering
    	\includegraphics[width=.9\textwidth]{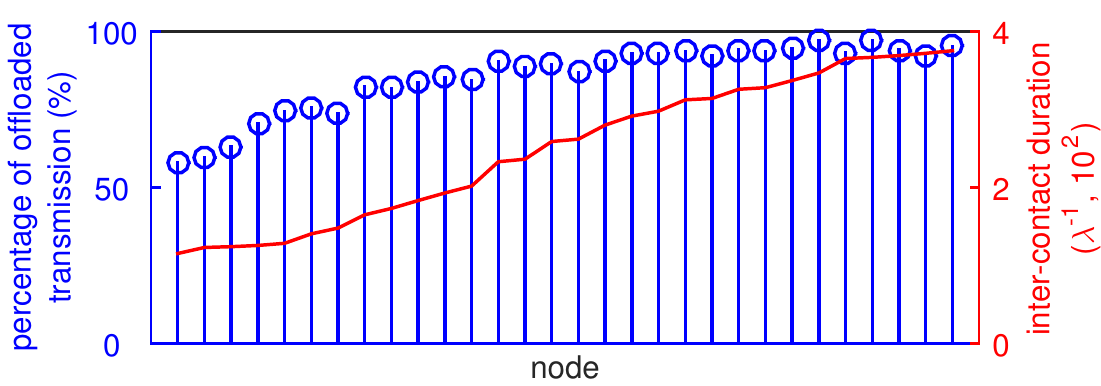}
    	\caption{expected inter-contact duration between node and infrastructure}
    	\label{fig:lambda}
    \end{subfigure}
    \\
    \vspace{0.1cm}
	\begin{subfigure}[t]{0.48\textwidth}
		\centering
    	\includegraphics[width=.9\textwidth]{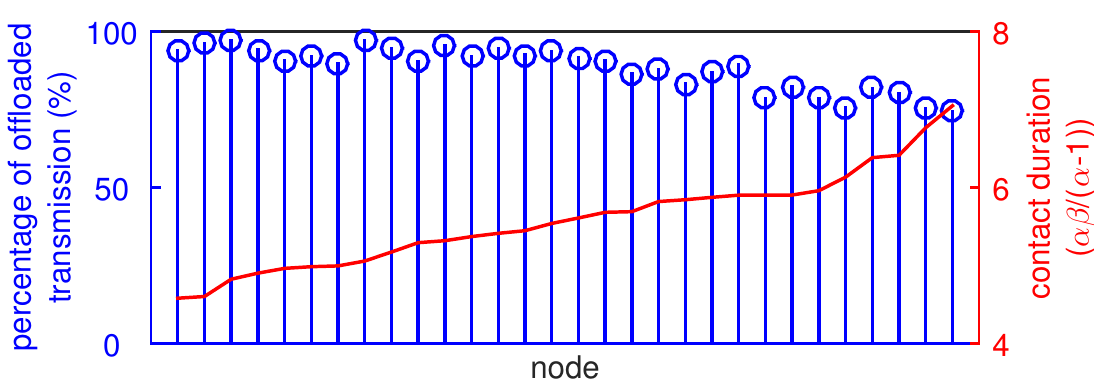}
    	\caption{expected contact duration between node and infrastructure}
    	\label{fig:alpha}
    \end{subfigure}    
\caption{Percentage of offloaded data transmissions at nodes with different inter-contact duration and contact duration to infrastructure, where $d=10$, $d_m=15$, $\lambda'=[0.001,0.01]$, $\alpha=[6,10]$, $S$ varies from 2 to 100, and $T$ varies from 20 to 1000.}
\label{fig:alpha&lambda}
\end{figure}

\begin{figure*}[t]
\setlength{\abovecaptionskip}{15pt}
\setlength{\belowcaptionskip}{-10pt}
\centering
	\begin{subfigure}[t]{0.3\textwidth}
		\centering
    	\includegraphics[width=.9\textwidth]{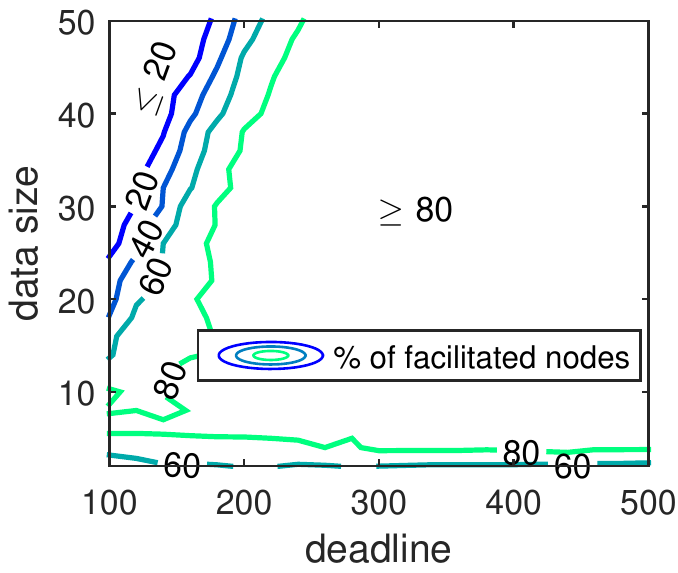}
    	\caption{$d=10$, $d_m=15$, $\alpha=[6,10]$, $\lambda'=[0.001,0.01]$}
    	\label{fig:contour1}
    \end{subfigure}
    \hspace{0.2cm}
    \begin{subfigure}[t]{0.3\textwidth}
    	\centering
    	\includegraphics[width=.9\textwidth]{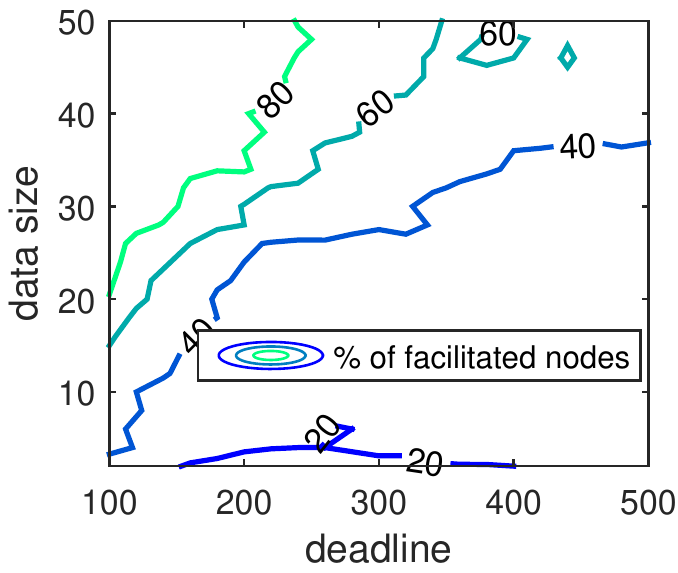}
    	\caption{$d=10$, $d_m=15$, $\alpha=[6,10]$, $\lambda'=[0.01,0.1]$}
    	\label{fig:contour2}
    \end{subfigure}
    \hspace{0.2cm}
    \begin{subfigure}[t]{0.3\textwidth}
    	\centering
    	\includegraphics[width=.9\textwidth]{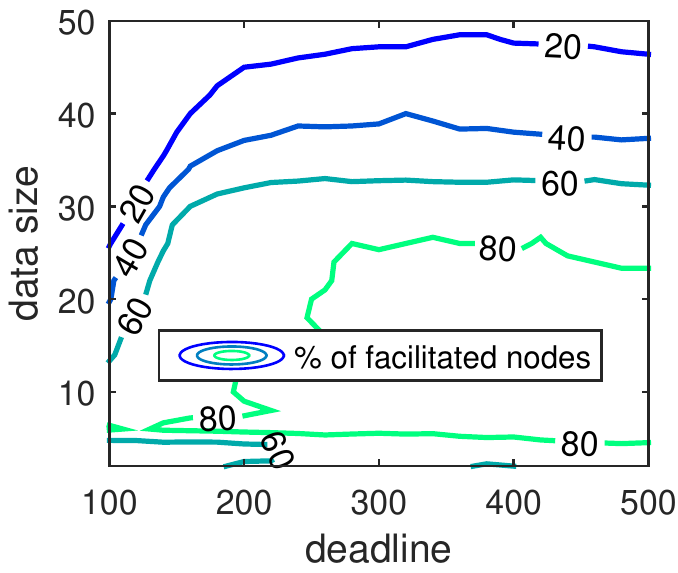}
    	\caption{$d=10$, $d_m=15$, $\alpha=[3,4]$, $\lambda'=[0.001,0.01]$}
    	\label{fig:contour3}
    \end{subfigure}
    \\
    \vspace*{0.5cm}
    \begin{subfigure}[t]{0.3\textwidth}
    	\centering
    	\includegraphics[width=.9\textwidth]{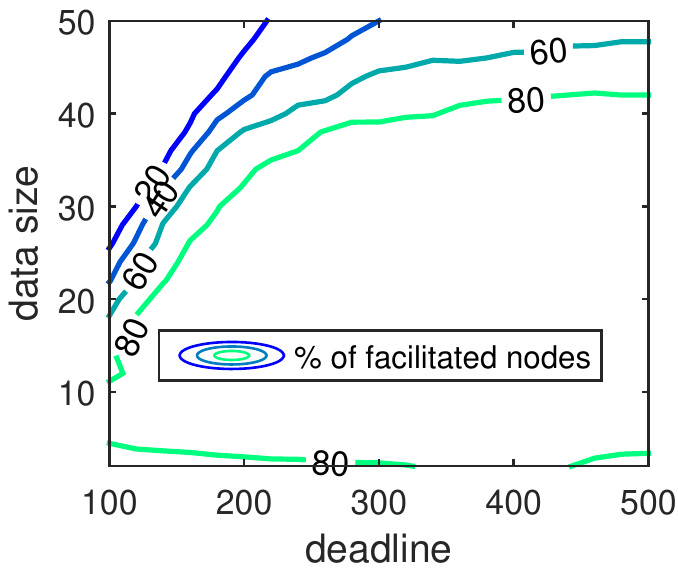}
    	\caption{$d=5$, $d_m=8$, $\alpha=[6,10]$, $\lambda'=[0.001,0.01]$}
    	\label{fig:contour4}
    \end{subfigure}
    \hspace{0.2cm}
    \begin{subfigure}[t]{0.3\textwidth}
    	\centering
    	\includegraphics[width=.9\textwidth]{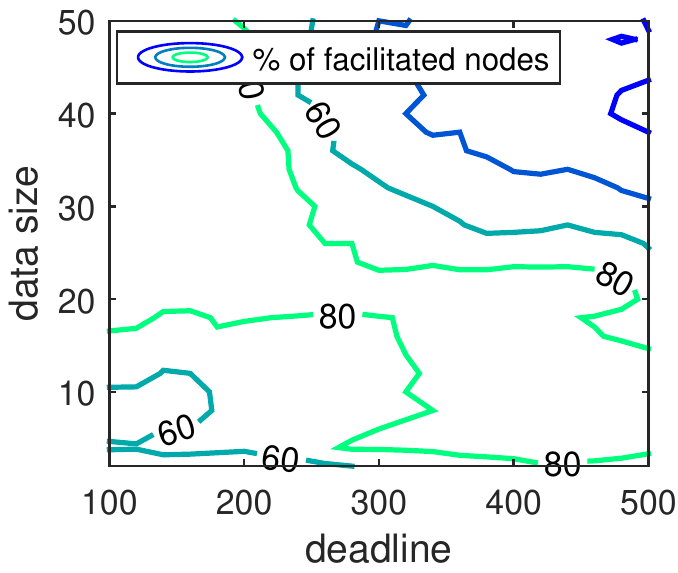}
    	\caption{$d=5$, $d_m=8$, $\alpha=[6,10]$, $\lambda'=[0.01,0.1]$}
    	\label{fig:contour5}
    \end{subfigure}
    \hspace{0.2cm}
    \begin{subfigure}[t]{0.3\textwidth}
    	\centering
    	\includegraphics[width=.9\textwidth]{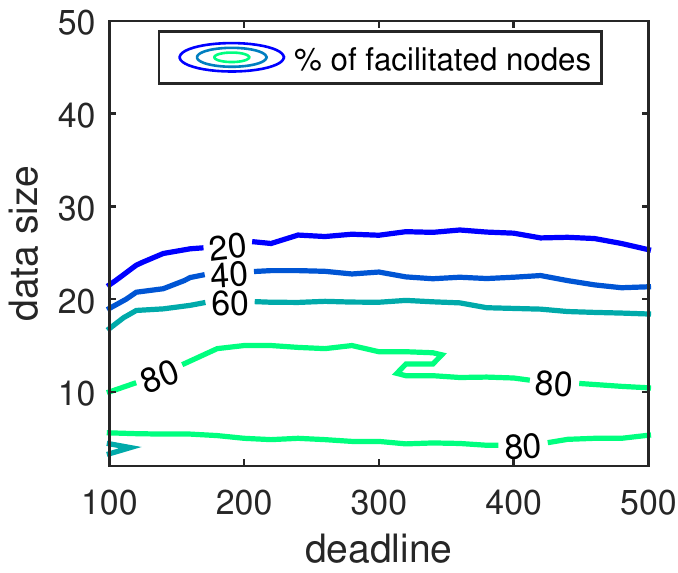}
    	\caption{$d=5$, $d_m=8$, $\alpha=[3,4]$, $\lambda'=[0.001,0.01]$}
    	\label{fig:contour6}
    \end{subfigure}
\caption{Percentage of facilitated nodes for data transmissions with different data sizes and deadlines in various network settings.}
\label{fig:contour}
\end{figure*}

\textbf{Delivery probability estimation.} As shown in Figure \ref{fig:datasize}, for \emph{Individual}, the estimated delivery probability and the simulated probability are almost the same for different data sizes. Meanwhile, for \emph{Cooperative} there is a little difference between the estimated probability and the simulated probability when the data size is large. That might be incurred by the approximation of sum of Pareto random variables as discussed in Section \ref{sec:framework}. When the data size increases, more node contacts are required to transfer a data item between neighboring nodes, which may increase the deviation of the approximation. However, this will not impact the decision of cooperative offloading, since the estimated probability of \emph{Cooperative} is still much higher than \emph{Individual}. As the deadline is irrelevant to the approximation, the increase of the deadline does not impact the difference between the estimated and simulated probabilities, and they are almost identical as shown in Figure \ref{fig:deadline}.

\textbf{\textit{Cooperative} is close to the optimum.}
We also compare \emph{Cooperative} to the optimal solution of the cooperative offloading problem, denoted as \emph{Optimal}. Due to the high computational complexity of \emph{Optimal}, we choose a problem of small size (i.e., we randomly select one node in a generated synthetic network to send data to infrastructure) and use brute-force search to find the optimum. 

Figures \ref{fig:datasizeopt} and \ref{fig:deadlineopt} depict the delivery probabilities of \emph{Cooperative} and \emph{Optimal} in terms of varying data size and deadline, respectively. \emph{Cooperative} is close to \emph{Optimal} in both cases and different data sizes and deadlines only slightly affect their disparities. 

\textbf{When data offloading benefits.} Figure \ref{fig:alpha&lambda} gives the relation between percentage of offloaded data transmissions (over all the data transmissions with different data sizes and deadlines) and contact pattern with infrastructure at particular nodes, where contact pattern is captured by expected inter-contact duration and contact duration between a node and infrastructure. Intuitively, nodes that frequently contact with infrastructure or have long contact duration with infrastructure do not need to offload the data as much as other nodes. Consistent with this intuition, as depicted in Figure \ref{fig:lambda}, the more frequently a node contacts infrastructure, the less transmissions are offloaded. Moreover, as shown in Figure \ref{fig:alpha}, the shorter the contact duration, the more transmissions are offloaded. Moreover,
inter-contact duration makes a larger impact on offloaded transmissions than contact duration. For example, when the contact frequency is low, a node may still need to offload the data even if the contact duration is long, unless the data is transmitted during the contact. That is the reason offloaded transmissions decrease slow with the increase of contact duration as shown in Figure \ref{fig:alpha}.

\textbf{Network settings affect data offloading.} Figure \ref{fig:contour} gives the contour of the percentage of facilitated nodes for data transmissions with different data sizes and deadlines in various network settings, where the facilitated node means that data transmission with particular data size and deadline is offloaded at the node. For example, the area enclosed by the green line (80\%) in Figure \ref{fig:contour1} indicates there are more than or equal to 80\% of nodes that offload the transmissions with corresponding data size and deadline.

As shown in Figure \ref{fig:contour1}, for the network with $a=[6,10]$ and $\lambda'=[0.001, 0.01]$, most of the transmissions are offloaded (i.e. most of area is enclosed by green line (80\%)) except the transmissions with large data sizes and short deadlines shown as the upper left corner. As shown in Figure \ref{fig:contour2}, when nodes more frequently contact infrastructure (i.e. $\lambda'=[0.01, 0.1]$), small data can be easily transmitted to infrastructure directly, and thus the transmissions with small data size are not frequently offloaded (less than 40\%). On the contrary, the transmissions with large data sizes and short deadlines are mostly offloaded. When contact duration between nodes becomes short (i.e. $\alpha=[3,4]$ as in Figure \ref{fig:contour3}), compared to Figure \ref{fig:contour1}, the delivery probability of the transmissions with large data sizes cannot be greatly improved and thus they are less offloaded at nodes. However, the transmissions with small data sizes are mostly offloaded.


\begin{figure}[!t]
\centering
	\begin{subfigure}[t]{0.48\textwidth}
		\centering
    	\includegraphics[width=.9\textwidth]{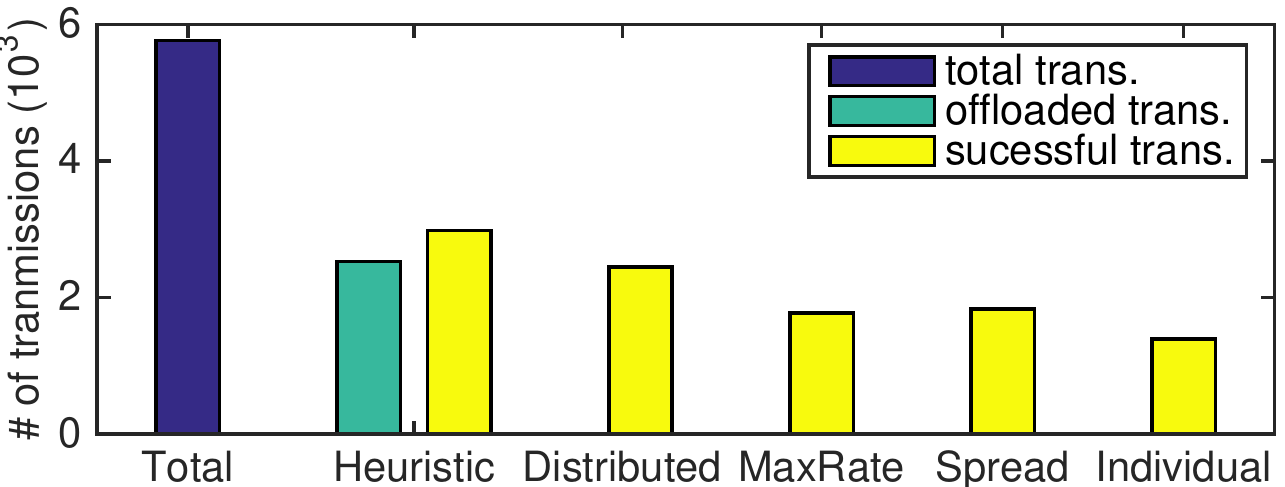}
    	\caption{MIT Reality, $S=[10,60]$ MB and $T=[10,100]$ hours}
    	\label{fig:mit}
    \end{subfigure}
    \\
    \vspace{0.1cm}
    \begin{subfigure}[t]{0.48\textwidth}
 	\centering    	
    	\includegraphics[width=.9\textwidth]{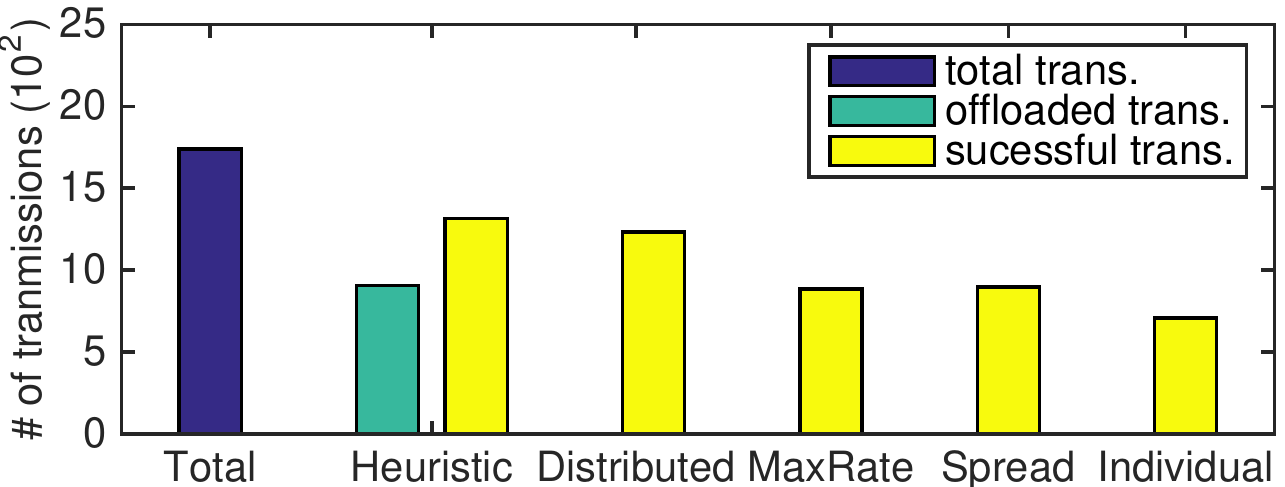}
    	\caption{DieselNet, $S=[2,10]$ MB and $T=[1,10]$ hours}
    	\label{fig:diesel}
    \end{subfigure}
\caption{Total transmissions, offloaded transmissions of \textit{Heuristic} and successful transmissions of \textit{Heuristic}, \textit{Distributed}, \textit{MaxRate}, \textit{Spread} and \textit{Individual} in MIT Reality and DieselNet.}
\label{fig:mit&diesel}
\end{figure}

When nodes have a small number of neighbors in a network, e.g. $d=5$ and $d_m=8$ as in Figure \ref{fig:contour4}, compared to Figure \ref{fig:contour1}, the reduced node degree undermines the offloading capability of the network; i.e., the offloaded transmissions are generally less than that of Figure \ref{fig:contour1}. Compared with Figure \ref{fig:contour4}, inter-contact duration between node and infrastructure is decreased ($\lambda'=[0.01,0.1]$) in Figure \ref{fig:contour5}. This dramatically changes the pattern of the offloaded transmissions. As shown in Figure \ref{fig:contour5}, offloading can improve the probability of most transmissions but not for the transmissions with large data sizes and long deadlines. That is because when nodes frequently contact infrastructure, it is easy for nodes to transmit a data item to infrastructure by multiple contacts and offloading the data item to other nodes may not yield a high delivery probability. Last, let us compare Figure \ref{fig:contour6} with Figure \ref{fig:contour3}. It is shown again that node degree affects the offloading capability of the network, since node degree determines how many nodes nearby that can be explored for data offloading.


Therefore, from Figure~\ref{fig:contour}, we can see network settings do affect data offloading. In general, when node degree is higher, there are more paths to infrastructure and thus data transmissions are more often offloaded. When nodes contact infrastructure more frequently, nodes can transmit more data directly to infrastructure and thus data transmissions are less often offloaded. When contact duration between nodes becomes shorter, offloading becomes less possible to increase data delivery probability and thus data transmissions are less often offloaded. 

In summary, through the extensive experiments on synthetic networks, we can conclude that the proposed probabilistic framework can accurately estimate data delivery probability over opportunistic paths, cooperative data offloading can significantly improve data delivery probability in various network settings and contact patterns, and the data delivery probability achieved by the heuristic algorithm is close to the optimum.

\subsection{Evaluations on Real Traces}
Next, we evaluate the performance of the heuristic algorithm and the distributed algorithm based on real traces. We compare them with other two solutions: \emph{Spread}, where nodes offload the carried data to any encountered node, and \emph{MaxRate}, where nodes only transmit the carried data to infrastructure or the node in its neighbor set that has the maximum contact rate with infrastructure. We also give the performance of no offloading, i.e., the node directly sends data to infrastructure, denoted as \emph{Individual}.

The two opportunistic mobile network traces used are MIT Reality \cite{eagle2006reality} and DieselNet \cite{balasubramanian2007dtn}. They record contacts among mobile devices equipped with Bluetooth or WiFi moving on university campus (MIT Reality) and in suburban area (DieselNet). The details of these two traces are summarized in Table \ref{tab:Traces}.

\begin{scriptsize}
\begin{table}[h]
\renewcommand{\arraystretch}{1.2}
\caption{Trace summary}
\label{tab:Traces}
\centering
\begin{tabular}{c|c|c}
\hline
\textbf{Trace}           & \textbf{MIT Reality} & \textbf{DieselNet} \\
\hline\hline
Network type    & Bluetooth   & WiFi      \\
Contact type    & Direct      & Direct    \\
No. of devices  & 97          & 40        \\
Duration(days)  & 246         & 20        \\
No. of contacts & 3268        & 114,046   \\
\hline
\end{tabular}
\end{table}
\end{scriptsize}

In the experiment, half of the trace is used as warmup to obtain the distribution information of contact frequency and contact duration between nodes, and other half is used to run data transmission. Since there is no infrastructure in either trace, we choose the node with the maximum degree to act as infrastructure. Nodes that cannot construct two-hop path or one-hop to infrastructure are excluded from the traces (note that only very few nodes are eliminated). In the simulation, nodes send data items with different sizes and deadlines to infrastructure at a randomly selected timestamp in each simulation run and the results are averaged for 50 runs.

For MIT Reality trace, due to the Bluetooth scan interval, only the contacts with duration of five minutes or more are recorded, where contact duration is at minute level and inter-contact duration is at day level. Thus, for data transmissions, the data size varies from 10MB to 60MB (the data rate is set to 240Kbps) and the time constraint varies from 10 to 100 hours. For DieselNet, contact duration is at second level and inter-contact duration is at hour level. Thus, for data transmissions in DieselNet, the data size varies from 2MB to 10MB (the data rate is set to 3.2Mbps) and the deadline varies from 1 to 10 hours.

Figures \ref{fig:mit} and \ref{fig:diesel} give the total transmissions, the offloaded transmission of \textit{Heuristic}, the successful transmissions of \textit{Heuristic}, \textit{Distributed}, \textit{MaxRate}, \textit{Spread} and \textit{Individual} in MIT Reality and DieselNet, respectively. In MIT Reality, as shown in Figure \ref{fig:mit}, there are totally near 6000 data transmissions. If no offloading (\textit{Individual}) is performed, the number of successful transmissions is only about 1400. However, \textit{Heuristic} has about 3000 successful transmissions which is more than two times of \textit{Individual}. This is also better than other three algorithms: about 20\% higher than that of \textit{Distributed} and 70\% higher than that of \textit{MaxRate} and \textit{Spread}. Moreover, about 40\% of total transmissions are offloaded by by \textit{Heuristic}. The number of successful transmissions are higher than the offloaded transmissions because the transmissions that are not offloaded might be completed via direct contact between source node and infrastructure. \textit{Distributed} is close to \textit{Heuristic}, better than \textit{MaxRate} and \textit{Spread}, and almost two times of \textit{Individual}. Although \textit{Spread} and \textit{MaxRate} have different offload strategies, their performance is similar. 

In DieselNet, as shown in Figure \ref{fig:diesel}, there are about 1800 total transmissions. The ratio of successful transmissions of \textit{Heuristic} and \textit{Distributed} are both high, more than  70\%. \textit{Heuristic} offloaded more than 50\% of total transmissions. The successful transmissions of \textit{Heuristic} and \textit{Distributed} are 40\% more than that of \textit{MaxRate} and \textit{Spread} (their performance is still similar in DieselNet, about 900) and almost two times of \textit{Individual}.

Based on the evaluations on the real traces, it can be concluded that the heuristic algorithm performs better than the distributed algorithm, both of them are much better than the simple offloading strategies \emph{MaxRate} and \emph{Spread}, and they also perform about two times better than no offloading in both traces. In spite of the lack of global information, the performance of the distributed algorithm that only exploits paths no more than two hops for data transmissions is comparable to that of the heuristic algorithm.

\section{Conclusion}
\label{sec:Conclusion}
In this paper, we addressed the problem of cooperatively offloading data among opportunistically connected mobile devices so as to improve the probability of data delivery to infrastructure. We first provided the probabilistic framework to estimate the probability of data delivery over the opportunistic path and then, based on that, we proposed a heuristic algorithm to solve cooperative offloading. To cope with the lack of global information, we further proposed a distributed algorithm. The evaluation results show that the probabilistic framework accurately estimates the data delivery probability, cooperative offloading greatly improves the delivery probability, the heuristic algorithm approximates the optimum, and the performance of the heuristic algorithm and distributed algorithm outperforms other approaches.

\appendices
\section{Approximating Sum of Pareto Variables}
\label{sec:appendix}

For a collection of \emph{i.i.d.} random variables $\{D_i,\,i=1,\ldots,c\}$ and $D_i \sim \mathrm{Pareto}(\alpha,\beta)$, let $\mathcal{D}=\sum_{i=1}^c D_i$, $\mathcal{M}$ be the maximum value of $\{D_i,\,i=1,\ldots,c\}$, and $R=\mathcal{D}/\mathcal{M}$. Then, $\mathcal{D}$ can be approximated by $\mathcal{M}$, considering the ratio $R$. First let us consider the ratio of $D_1/\mathcal{M}$. Then the Cumulative Density Function (CDF) of $D_1/\mathcal{M}$ can be expressed in terms of the conditional distribution $F(x|y)$ of $D_1$ on the fixed maximum $\mathcal{M}=y$ \cite{Zaliapin2005approx} as following
\begin{equation}
\nonumber\begin{split}
G(z) =P(D_1/\mathcal{M}<z) &= \int_{0}^{\infty} P(\frac{D_1}{y}<z|y)dF^c(y) \\
                              &= \int_{0}^{\infty} F(yz|y)dF^c(y).
\end{split}
\end{equation}
By taking the derivative with respect to $z$, then
\begin{equation}
\label{eq:gz}
g(z) = \int_{\beta}^{\infty} yf(yz|y)dF^c(y).
\end{equation}
The conditional density $f(x|y)$ corresponding to the distribution $F(x|y)$ is given by
\begin{equation}
\label{eq:fxy}
f(x|y)=\frac{\delta(y-x)}{c}+(1-\frac{1}{c})\frac{f(x)H(y-x)}{F(y)},
\end{equation}
where $H(x) = 1$ for $x>0$, otherwise $H(x)=0$. The first term on the rhs of \eqref{eq:fxy} corresponds to the case $\mathcal{M}=D_1$, whose probability is $1/c$, while the complementary event $D_1 < \mathcal{M}$
occurs with probability $(1-1/c)$. By plugging \eqref{eq:fxy} into \eqref{eq:gz}, we then have
\begin{equation}
\nonumber\begin{split}
g(z) &= \int_{0}^{\infty}\left(\frac{\delta(1-z)}{c}+\frac{c-1}{c}\cdot\frac{f(yz)H(y-yz)}{F(y)}\right)dF^c(y) \\
        &=\frac{\delta(1-z)}{c}+(c-1)\int_{0}^{\infty}yf(yz)f(y)F^{c-2}(y)dy.
\end{split}
\end{equation}
Then, the expectation of the ratio $R=\mathcal{D}/\mathcal{M}$, denoted as $\bar{R}$, can be calculated as
\begin{equation}
\nonumber\begin{split}
\bar{R} &= c \int_0^1zg(z)dz \\
               &= 1+c(c-1) \cdot \int_{0}^{\infty} yf(y)F^{c-2}(y)\int_0^1 zf(yz)dydz \\
               &= 1+c(c-1) \cdot \\
               &\quad\quad\quad\quad \int_{0}^{\infty} yf(y)F^{c-2}(y) \left( \frac{F(y)}{y} - \frac{1}{y^2} \int_{0}^{y} F(x)dx \right) dy \\
               &= 1+(c-1)\left( \int_{0}^{\infty} dF^c(y) \right) -\\
               &\quad\quad\quad\quad c \left(\int_{0}^{\infty} \left( \frac{1}{y} \int_{0}^y F(u)du \right) dF^{c-1}(y) \right) \\
               &= c \left( 1- \int_{0}^{\infty} \left( \frac{1}{y} \int_{0}^y F(u)du \right) dF^{c-1}(y) \right).
\end{split}
\end{equation}
For Pareto distribution $F(x)=1-(\frac{\beta}{x})^{\alpha}$, then
\begin{equation}
\nonumber\begin{split}
\bar{R}  &=  c - c\int_{\beta}^{\infty}  1 - \frac{(\frac{\beta}{y})^{\alpha}-\frac{\alpha\beta}{y}}{1-a} d\left( 1- (\frac{\beta}{y})^\alpha\right)^{c-1}. \\
\end{split}
\end{equation}
By replacing the integral by \textit{beta function} $B(\cdot,\cdot)$, finally we have
\begin{equation}
\label{eq:R}
\bar{R}=\left\{\begin{matrix}
\frac{1-cB(c,\alpha^{-1})}{1-\alpha}, & \alpha \neq 1\\
\overset{c}{\underset{i=1}{\sum}}\,i^{-1}, & \alpha=1.
\end{matrix}\right.
\end{equation}
Thus, $P(\mathcal{D} \geq D)$ can be approximated as
\begin{equation}
\nonumber\begin{split}
P(\mathcal{D} \geq D) &\approx P(\mathcal{M}\bar{R} \geq D) = P(\mathcal{M} \geq \frac{D}{\bar{R}}) \\
&= 1- \prod_{i=1}^c P(D_i < \frac{D}{\bar{R}}) = 1 - \left(1 - (\frac{\beta\bar{R}}{D})^\alpha\right)^c, 
\end{split}
\end{equation}
where $\bar{R}$ can be easily calculated using \eqref{eq:R}.

\section*{Acknowledgment}
This work was supported in part by Network Science CTA under grant W911NF-09-2-0053 and CERDEC via contract W911NF-09-D-0006. A preliminary version of this work appeared in the Proceedings of IEEE INFOCOM 2016 \cite{lu2016cooperative}.

\bibliographystyle{abbrv}
\bibliography{ref}

\begin{thebibliography}{10}

\bibitem{dtn2}
{DTN2}.
\newblock \url{http://www.dtnrg.org}.

\bibitem{balasubramanian2007dtn}
A.~Balasubramanian, B.~Levine, and A.~Venkataramani.
\newblock Dtn routing as a resource allocation problem.
\newblock In {\em Proc. ACM SIGCOMM}, 2007.

\bibitem{banerjee2007energy}
N.~Banerjee, M.~D. Corner, and B.~N. Levine.
\newblock An energy-efficient architecture for dtn throwboxes.
\newblock In {\em Proc. IEEE NFOCOM}, 2007.

\bibitem{chun2011clonecloud}
B.~Chun, S.~Ihm, P.~Maniatis, M.~Naik, and A.~Patti.
\newblock Clonecloud: elastic execution between mobile device and cloud.
\newblock In {\em Proc. ACM EuroSys}, 2011.

\bibitem{cuervo2010maui}
E.~Cuervo, A.~Balasubramanian, D.-k. Cho, A.~Wolman, S.~Saroiu, R.~Chandra, and
  P.~Bahl.
\newblock Maui: making smartphones last longer with code offload.
\newblock In {\em Proc. ACM MobiSys}, 2010.

\bibitem{daly2007social}
E.~Daly and M.~Haahr.
\newblock Social network analysis for routing in disconnected delay-tolerant
  manets.
\newblock In {\em Proc. ACM MobiHoc}, 2007.

\bibitem{eagle2006reality}
N.~Eagle and A.~Pentland.
\newblock Reality mining: sensing complex social systems.
\newblock {\em Personal and Ubiquitous Computing}, 10(4):255--268, 2006.

\bibitem{erramilli2007diversity}
V.~Erramilli, A.~Chaintreau, M.~Crovella, and C.~Diot.
\newblock Diversity of forwarding paths in pocket switched networks.
\newblock In {\em Proc. ACM IMC}, 2007.

\bibitem{erramilli2008delegation}
V.~Erramilli, M.~Crovella, A.~Chaintreau, and C.~Diot.
\newblock Delegation forwarding.
\newblock In {\em Proc. ACM MobiHoc}, 2008.

\bibitem{gao2012social}
W.~Gao, Q.~Li, B.~Zhao, and G.~Cao.
\newblock Social-aware multicast in disruption-tolerant networks.
\newblock {\em Networking, IEEE/ACM Transactions on}, 20(5):1553--1566, 2012.

\bibitem{gilbert2011peer}
P.~Gilbert, V.~Ramasubramanian, P.~Stuedi, and D.~Terry.
\newblock Peer-to-peer data replication meets delay tolerant networking.
\newblock In {\em Proc. IEEE ICDCS}, 2011.

\bibitem{han2012mobile}
B.~Han, P.~Hui, V.~A. Kumar, M.~V. Marathe, J.~Shao, and A.~Srinivasan.
\newblock Mobile data offloading through opportunistic communications and
  social participation.
\newblock {\em IEEE Transactions on Mobile Computing}, 11(5):821--834, 2012.

\bibitem{hossmann2010know}
T.~Hossmann, T.~Spyropoulos, and F.~Legendre.
\newblock Know thy neighbor: Towards optimal mapping of contacts to social
  graphs for dtn routing.
\newblock In {\em Proc. IEEE INFOCOM}, 2010.

\bibitem{jain2005using}
S.~Jain, M.~Demmer, R.~Patra, and K.~Fall.
\newblock Using redundancy to cope with failures in a delay tolerant network.
\newblock In {\em Proc. ACM SIGCOMM}, 2005.

\bibitem{karagiannis2010power}
T.~Karagiannis, J.-Y. Le~Boudec, and M.~Vojnovic.
\newblock Power law and exponential decay of intercontact times between mobile
  devices.
\newblock {\em IEEE Transactions on Mobile Computing}, 9(10):1377--1390, 2010.

\bibitem{lancichinetti2009benchmarks}
A.~Lancichinetti and S.~Fortunato.
\newblock Benchmarks for testing community detection algorithms on directed and
  weighted graphs with overlapping communities.
\newblock {\em Physical Review E}, 80(1):016118, 2009.

\bibitem{li2014can}
Y.~Li and W.~Wang.
\newblock Can mobile cloudlets support mobile applications?
\newblock In {\em Proc. IEEE INFOCOM}, 2014.

\bibitem{lin2008efficient}
Y.~Lin, B.~Li, and B.~Liang.
\newblock Efficient network coded data transmissions in disruption tolerant
  networks.
\newblock In {\em Proc. IEEE INFOCOM}, 2008.

\bibitem{lin2008stochastic}
Y.~Lin, B.~Li, and B.~Liang.
\newblock Stochastic analysis of network coding in epidemic routing.
\newblock {\em Selected Areas in Communications, IEEE Journal on},
  26(5):794--808, 2008.

\bibitem{liu2013exploring}
S.~Liu and A.~D. Striegel.
\newblock Exploring the potential in practice for opportunistic networks
  amongst smart mobile devices.
\newblock In {\em Proc. ACM MobiCom}, 2013.

\bibitem{lu2016networking}
Z.~Lu, G.~Cao, and T.~La~Porta.
\newblock Networking smartphones for disaster recovery.
\newblock In {\em Proc. IEEE PerCom}, 2016.

\bibitem{lu2016cooperative}
Z.~Lu, X.~Sun, and T.~La~Porta.
\newblock Cooperative data offloading in opportunistic mobile networks.
\newblock In {\em Proc. IEEE INFOCOM}, 2016.

\bibitem{lu2014skeleton}
Z.~Lu, X.~Sun, Y.~Wen, and G.~Cao.
\newblock Skeleton construction in mobile social networks: Algorithms and
  applications.
\newblock In {\em Proc. IEEE SECON}, 2014.

\bibitem{lu2015algorithms}
Z.~Lu, X.~Sun, Y.~Wen, G.~Cao, and T.~La~Porta.
\newblock Algorithms and applications for community detection in weighted
  networks.
\newblock {\em IEEE Transactions on Parallel and Distributed Systems},
  26(11):2916--2926, 2015.

\bibitem{lu2014information}
Z.~Lu, Y.~Wen, and G.~Cao.
\newblock Information diffusion in mobile social networks: The speed
  perspective.
\newblock In {\em Proc. IEEE INFOCOM}, 2014.

\bibitem{pietilanen2012dissemination}
A.-K. Pietil{\"a}nen and C.~Diot.
\newblock Dissemination in opportunistic social networks: the role of temporal
  communities.
\newblock In {\em ACM MobiHoc}, 2012.

\bibitem{sciancalepore2016offloading}
V.~Sciancalepore, D.~Giustiniano, A.~Banchs, and A.~Hossmann-Picu.
\newblock Offloading cellular traffic through opportunistic communications:
  Analysis and optimization.
\newblock {\em IEEE Journal on Selected Areas in Communications},
  34(1):122--137, 2016.

\bibitem{sermpezis2014not}
P.~Sermpezis and T.~Spyropoulos.
\newblock Not all content is created equal: effect of popularity and
  availability for content-centric opportunistic networking.
\newblock In {\em Proc. ACM MobiHoc}, 2014.

\bibitem{shi2012serendipity}
C.~Shi, V.~Lakafosis, M.~H. Ammar, and E.~W. Zegura.
\newblock Serendipity: enabling remote computing among intermittently connected
  mobile devices.
\newblock In {\em Proc. ACM MobiHoc}, 2012.

\bibitem{spyropoulos2005spray}
T.~Spyropoulos, K.~Psounis, and C.~S. Raghavendra.
\newblock Spray and wait: an efficient routing scheme for intermittently
  connected mobile networks.
\newblock In {\em Proc. ACM WDTN}, 2005.

\bibitem{spyropoulos2008efficient}
T.~Spyropoulos, K.~Psounis, and C.~S. Raghavendra.
\newblock Efficient routing in intermittently connected mobile networks: the
  multiple-copy case.
\newblock {\em IEEE/ACM Transactions on Networking}, 16(1):77--90, 2008.

\bibitem{wang2014tempo}
S.~Wang, X.~Wang, X.~Cheng, J.~Huang, and R.~Bie.
\newblock The tempo-spatial information dissemination properties of mobile
  opportunistic networks with levy mobility.
\newblock In {\em Proc. IEEE ICDCS}, 2014.

\bibitem{wang2014toss}
X.~Wang, M.~Chen, Z.~Han, D.~O. Wu, and T.~T. Kwon.
\newblock Toss: Traffic offloading by social network service-based
  opportunistic sharing in mobile social networks.
\newblock In {\em Proc. IEEE INFOCOM}, 2014.

\bibitem{xiang2014ready}
L.~Xiang, S.~Ye, Y.~Feng, B.~Li, and B.~Li.
\newblock Ready, set, go: Coalesced offloading from mobile devices to the
  cloud.
\newblock In {\em Proc. IEEE INFOCOM}, 2014.

\bibitem{yuan2009predict}
Q.~Yuan, I.~Cardei, and J.~Wu.
\newblock Predict and relay: an efficient routing in disruption-tolerant
  networks.
\newblock In {\em Proc. ACM MobiHoc}, 2009.

\bibitem{Zaliapin2005approx}
I.~Zaliapin, Y.~Kagan, and F.~Schoenberg.
\newblock Approximating the distribution of {P}areto sums.
\newblock {\em Pure and Applied geophysics}, 162(6-7):1187--1228, 2005.

\bibitem{zhang2015collaborative}
W.~Zhang, Y.~Wen, and D.~O. Wu.
\newblock Collaborative task execution in mobile cloud computing under a
  stochastic wireless channel.
\newblock {\em IEEE Transactions on Wireless Communications}, 14(1):81--93,
  2015.

\bibitem{zhang2013transient}
X.~Zhang and G.~Cao.
\newblock Transient community detection and its application to data forwarding
  in delay tolerant networks.
\newblock In {\em Proc. IEEE ICNP}, 2013.

\bibitem{zhuo2014incentive}
X.~Zhuo, W.~Gao, G.~Cao, and S.~Hua.
\newblock An incentive framework for cellular traffic offloading.
\newblock {\em IEEE Transactions on Mobile Computing}, 13(3):541--555, 2014.

\bibitem{zhuo2011contact}
X.~Zhuo, Q.~Li, W.~Gao, G.~Cao, and Y.~Dai.
\newblock Contact duration aware data replication in delay tolerant networks.
\newblock In {\em Proc. IEEE ICNP}, 2011.

\end{thebibliography}

\end{document}